\include{Commands}

\documentclass[11pt, letterpaper]{article}
\usepackage{geometry} 

\usepackage[utf8]{inputenc}

\usepackage[english]{babel}
\usepackage{amssymb, amsfonts, amsthm, amsmath, bm, graphicx}
\usepackage[authoryear]{natbib}
\usepackage[dvipsnames]{xcolor}
\usepackage{parskip}
\usepackage{tabu}
\usepackage{verbatim} 
\usepackage{helvet,times}
\usepackage{afterpage} 

\usepackage{titling}

\usepackage{newfloat}
\DeclareFloatingEnvironment[name={Supp. Figure}]{suppfigure}

\graphicspath{{../Figures/}}
\newcommand{\ord}{\operatorname{ord}}

\newcommand{\hf}{\fontfamily{phv}\selectfont}{}  

\title{ {\bf \hf
The role of traction in membrane curvature generation}}

\author{\bf \hf Haleh Alimohamadi$^1$, Ritvik Vasan$^1$, Julian E. Hassinger$^2$\\ \bf \hf Jeanne C. Stachowiak$^3$, Padmini Rangamani$^{1*}$}
\date{\small \hf
$^1$Department of Mechanical and Aerospace Engineering, \\
University of California San Diego, La Jolla CA 92093\\
$^2$Biophysics Graduate Program, University of California, Berkeley, CA 94720\\
$^3$Department of Biomedical Engineering, University of Texas at Austin, Austin, TX 78712\\
\medskip 
$^*$To whom correspondence should be addressed. e-mail: prangamani@ucsd.edu \\
\medskip 
\today}
 
\begin{document}

\maketitle


\textbf{Abstract} Curvature in biological membranes can be generated by a variety of
different molecular mechanisms such as protein scaffolding, lipid or
protein asymmetry, cytoskeletal forces, etc. These mechanisms have
the net effect of generating stresses on the bilayer that are translated
into distinct final shapes of the membrane. We propose reversing this input-output relationship by using the shape of a curved membrane to infer physical quantities like the magnitude of the applied forces acting on the bilayer. To do this, we calculate the normal and tangential tractions along the membrane using the known material properties of the membrane along with its shape. These tractions are a quantitative measure of the response of the membrane to external forces or sources of spontaneous curvature. We demonstrate the utility of this approach first by showing that the magnitude of applied force can be inferred from the shape of the membrane alone in both simulations and experiments of membrane tubulation. Next, we show that membrane budding by local differences in spontaneous curvature is driven purely by the generation of traction in the radial direction and the emergence of an effective line tension at the boundary of these regions. Finally, we show that performing this calculation on images of phase-separated giant vesicles yields a line tension similar to experimentally determined values.


\section*{Introduction}
\begin{figure}[t!]
\centering
\includegraphics[width=5in]{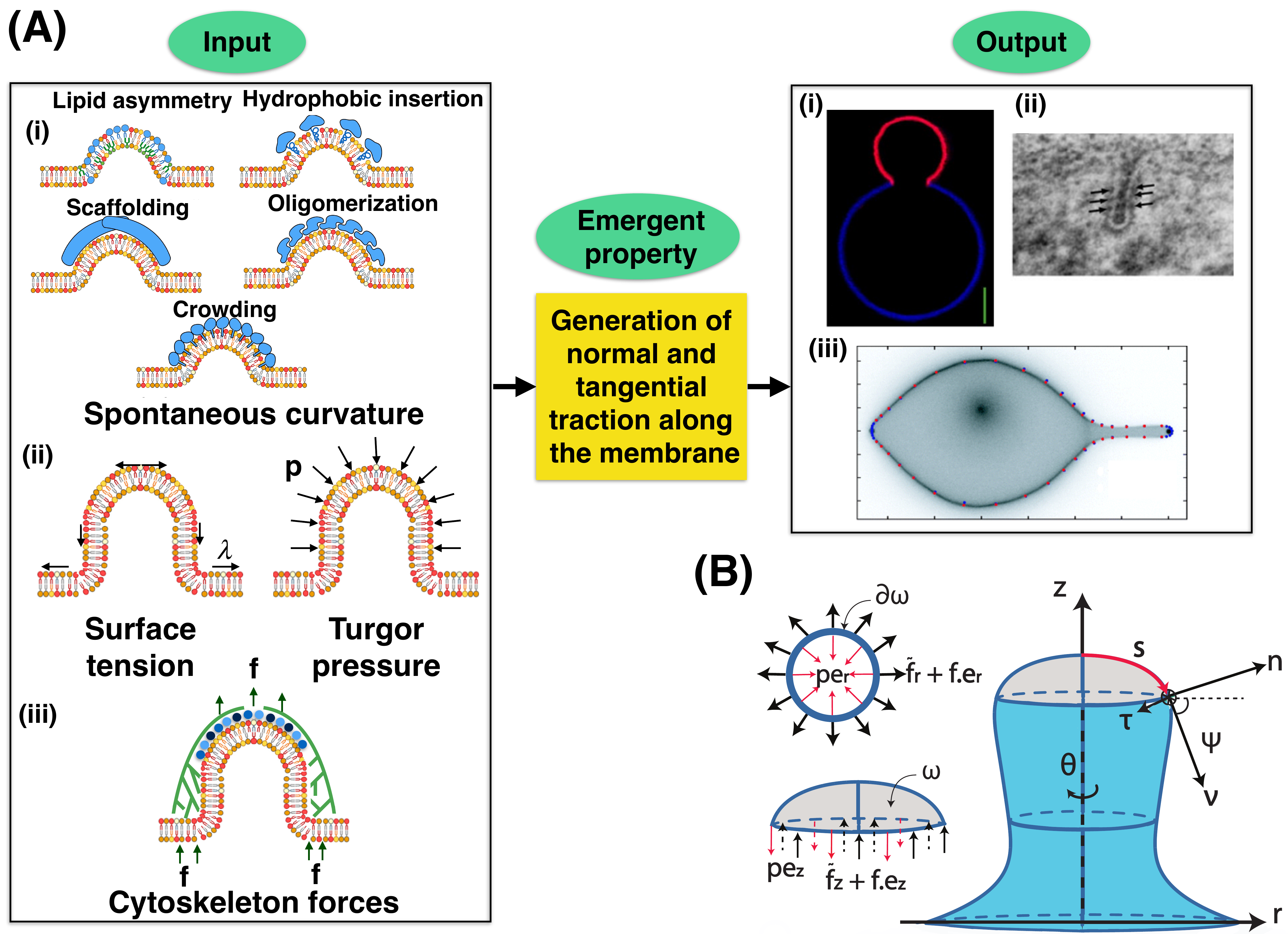}
\caption{Membrane curvature generation as an input-output relationship. (A) Membrane curvature is controlled by different physical inputs including (i) protein-induced spontaneous curvature, (ii) turgor pressure, and membrane tension, and (iii) forces exerted by the cytoskeleton. These seemingly different mechanisms exert normal and tangential tractions on the membrane that result in an output shape -- (i) observed in a two-photon microscopy image of an axially symmetric vesicle with fluid phase coexistence \cite{baumgart2005membrane}, (ii) a tubular endocytic invagination visualized by electron microscopy \cite{buser2013ultrastructural}, and (iii) a fluorescence image of a vesicle from a force-extension experiment \cite{lee2008membrane}. (B) Coordinate system for axisymmetric calculations of membrane shape and tractions. The $z$-axis is the axis of symmetry and $s$ is the arc length along the membrane. Rotation of the curve along the axis of symmetry generates the surface of revolution. At any given point, the tractions along the membrane can be calculated as $\tilde{f}_z$ and $\tilde{f}_r$ Eqs. (\ref{eq:s-traction-radial}, \ref{eq:s-traction-axial}). Inset shows that pressure opposes traction and external force in both radial and axial directions.}
\centering
\label{fig:schematic}
\end{figure}

Cell shape regulates function in development, differentiation, motility, and signal transduction \cite{neves2008cell,rangamani2013decoding,rangamani2011signaling,xiong2010mechanisms} and is exquisitely modulated by a large protein-cytoskeletal assembly with great precision \cite{rangamani2011signaling}. A centerpiece of cell shape regulation is the ability of cellular membranes to bend and curve; this is critical for a variety of cellular functions including membrane trafficking processes, cytokinetic abscission, and filopodial extension \cite{mukherjee2000role,mattila2008filopodia,shillcock2006computational}. In order to carry out these functions, cells harness diverse mechanisms of curvature generation including compositional heterogeneity \cite{baumgart2003imaging,baumgart2005membrane}, protein scaffolding \cite{karotki2011eisosome}, insertion of amphipathic helices into the bilayer \cite{lee2005sar1p, campelo2008hydrophobic}, and forces exerted by the cytoskeleton \cite{giardini2003compression}. Even reconstituted and synthetic membrane systems exhibit a wide range of shapes in response to different curvature-inducing mechanisms including steric pressure due to intrinsically disordered proteins \cite{busch2015intrinsically} and protein crowding \cite{stachowiak2012membrane,snead2017membrane}. These effects can be interpreted as input-output relationships, where the input is the protein distribution, lipid asymmetry, or the forces exerted by the cytoskeleton and the output is the observed shape of the membrane (Fig. \ref{fig:schematic}A).

Two previous studies motivated our desire to examine whether this input-output relationship could be reversed - using membrane shapes to gain insight about mechanisms of membrane curvature generation. Lee et al. suggested that membrane shape in and of itself is a reporter of applied forces \cite{lee2008membrane}. In this elegant study, the authors showed that calculating the axial force along the membrane based on its shape alone is sufficient to extract the magnitude of the applied force required to form a tether on a giant unilamellar vesicle (GUV). In the case of heterogeneous phase-separated membranes, multiple studies have shown that line tension is sufficient to induce membrane budding \cite{baumgart2005membrane,baumgart2003imaging}, and also it can cause scission during clathrin-mediated endocytosis \cite{liu2006endocytic}. These two studies used the principle of force balance to extract physical quantities from observed shapes of the membrane, an approach we seek to generalize here.

In this work, we have sought to answer the question, what information does the observed shape of the membrane contain? We hypothesized that the different input-output relationships for membrane curvature generation are connected by the normal and tangential tractions along the membrane. That is, given a membrane shape and some information about its composition, we can extract the tractions acting along the membrane. Furthermore, understanding how much stress needs to be applied on the membrane to curve it to a desired shape will give us sufficient information to potentially engineer novel mechanisms that can generate those stresses. To test this hypothesis, we used the Helfrich model of lipid bilayers \cite{helfrich1973elastic} to derive the equations of normal and tangential tractions acting on the membrane. We then applied our model to two classic membrane deformations -- tethers and buds -- and showed that our predictions of traction distributions match experimental measurements.

\section*{Model Development}
\subsection*{Assumptions}

We assume that the curvature of the membrane is much larger than the thickness of the bilayer and that the lipid bilayer can be modeled as a thin elastic shell using the Helfrich-Canham energy model \cite{helfrich1973elastic}. We also assume that the membrane is incompressible because the energetic cost of stretching the membrane is high \cite{rawicz2000effect}; this constraint is implemented using a Lagrange multiplier as discussed in \cite{rangamani2013interaction,jenkins1977static}. We assume that bending and Gaussian moduli are uniform throughout the membrane. Finally, for simplicity in the numerical simulations, we assume that the membrane in the region of interest is rotationally symmetric (Fig. \ref{fig:schematic}B). 


\subsection*{Helfrich energy and equations of motion}
We use a modified version of the Helfrich energy that includes spatially-varying spontaneous curvature $C \left(\theta^{\alpha} \right)$, \cite{steigmann1999fluid,hassinger2017design,agrawal2009boundary,rangamani2014protein},

\begin{align}
W = \kappa \left[ H - C(\theta^{\alpha})\right]^2 + \kappa_G K.
\label{eq:Helfrich}
\end{align}

\noindent where $W$ is the energy per unit area, $H$ is the local mean curvature, and $K$ is the local Gaussian curvature. $\theta^{\alpha}$ denotes the surface coordinates. This form of the energy density accommodates the local heterogeneity in the spontaneous curvature $C$ and differs from the standard Helfrich energy by a factor of $2$. Consequently, our bending modulus, $\kappa$, is twice that of the standard bending modulus typically  encountered in the literature. The used notation is given in Table \ref{table:notation}.

\begin{table}
\begin{center}
\caption{Notation used in the model}
\begin{tabular} {l l l }
\hline\hline
Notation &  Description & Units\\ [0.5ex]
\hline
$W$ & Local energy per unit area &  $\mathrm{pN/nm}$\\
$p$ & Pressure difference across the membrane  & $\mathrm{pN/nm^2}$ \\
$C$ & Spontaneous curvature & $\mathrm{nm^{-1}}$ \\
$\theta^{\alpha}$ & Parameters describing the surface\\  
${\bf r}$ & Position vector on the surface & \\
${\bf n}$ & Normal to the membrane surface &  unit vector\\
$\textbf{a}_\alpha$ & Basis vectors describing the tangent plane\\
$\lambda$ & Membrane tension, $-(W+\gamma)$  & $\mathrm{pN/nm}$\\
$H$ & Mean curvature of the membrane & $\mathrm{nm^{-1}}$\\
$K$ & Gaussian curvature of the membrane & $\mathrm{nm^{-2}}$\\
$\kappa$ & Bending modulus  & $\mathrm{pN \cdot nm}$\\
$\kappa_G$ & Gaussian modulus & $\mathrm{pN \cdot nm}$\\
$s$ & Arc length  &  $\mathrm{nm}$\\
$\psi$ &  Angle between $ \textbf{e}_{r}$  and  $\textbf{a}_{s}$ & \\
${\bf f}$ & Applied force per unit area &  $\mathrm{pN /nm^2}$\\
$\tilde{f}$ & Traction & $\mathrm{pN/nm}$\\
$\tilde{f_{n}}$ & Component of the traction in the normal direction & $\mathrm{pN/nm}$\\
$\tilde{f_{\nu}}$ & Component of the traction in the tangential direction & $\mathrm{pN/nm}$\\
$\tilde{F}_z$ &  Calculated force in axial direction  &  $\mathrm{pN}$ \\
$\xi$ &  Energy per unit length  &  $\mathrm{pN}$ \\
\hline
\label{table:notation}
\end{tabular}
\end{center}
\end{table}

A balance of stresses normal to the membrane yields the so-called ``shape equation'' for energy functional (Eq. \ref{eq:Helfrich}) {,} 

\begin{equation}
\underbrace{\Delta \left[\kappa \left(H - C \right)\right] + 2 \kappa \left( H - C \right) \left(2 H^2 - K \right) - 2 \kappa H \left( H - C \right)^2}_{\text{Elastic Effects}}
= \underbrace{p + 2 \lambda H}_{\text{\shortstack{Capillary \\ effect}}} + \underbrace{\textbf{f} \cdot \textbf{n}}_{\text{\shortstack{External \\ force}}}.
\label{eq:shape}
\end{equation}

\noindent where $\Delta$ is the surface Laplacian, $p$ is the pressure difference across the membrane, $\lambda$ is interpreted to be the membrane tension \cite{rangamani2014protein,steigmann1999fluid},  $\mathbf{f}$ is a force per unit area applied to the membrane surface, and $\mathbf{n}$ is the unit normal to the surface \cite{agrawal2009boundary, walani2015endocytic}. In this model, $\textbf{f}$ represents the applied force by the actin cytoskeleton, tether, or by any surface in contact with membrane. This force need not necessarily be normal to the membrane \cite{hassinger2017design}. 

A consequence of heterogenous protein-induced spontaneous curvature, heterogeneous moduli, and externally applied force is that $\lambda$ is not homogeneous along the membrane \cite{agrawal2009boundary,rangamani2014protein,hassinger2017design}. A balance of forces tangent to the membrane yields the spatial variation of membrane tension,
\begin{align}
\underbrace{\lambda_{, \alpha}}_{\text{\shortstack{Gradient of \\ surface tension}}}=  \underbrace{2 \kappa \left( H - C \right) \frac{\partial{C}}{\partial{{\theta}^{\alpha}}}}_{\text{protein-induced variation}} - \underbrace{\textbf{f}\cdot\textbf{a}_{\alpha}}_{\text{\shortstack{External\\ force}}}.
\label{eq:lambda}
\end{align}

\noindent where $\left(\cdot\right)_{, \alpha}$ is the partial derivative with respect to the coordinate $\alpha$ and $\textbf{a}_{\alpha}$ is the unit tangent in the $\alpha$ direction. $\lambda$ can be interpreted as the membrane tension \cite{rangamani2014protein,steigmann1999fluid}, and is affected by the spatial variation in spontaneous curvature and by the tangential components ($\textbf{a}_{\alpha}$) of any external  force. A complete derivation of the stress balance and the governing equations of motion, including the effect of variable bending and Gaussian moduli, is presented in the Supplementary online material (SOM).

\subsection*{Force balance along the membrane and traction}


We define the force balance along a surface $\omega$ bounded by a parallel line of constant $\theta^\alpha$ and the bounding curve denoted by $\partial \omega$ as

\begin{equation}
\underbrace{\int_{\omega} p\textbf{n} da}_{\text{\shortstack{Force due to pressure \\ acting on the surface}}}  {+} \underbrace{\int_{\partial \omega}\tilde{\textbf{f} }dt}_{\text{\shortstack{Traction force \\ along the boundary}}} =0,
\label{eq:force balance}
\end{equation}

\noindent where $\tilde{\textbf{f}}$ represents the traction along a curve bounded by $t$. These tractions give us information about the response of the membrane to external factors like applied loading or a protein coat. While Eq. (\ref{eq:force balance}) is general and independent of coordinates, we will restrict further analysis to axisymmetric coordinates, parametrized by arc length $s$ and azimuthal angle $\theta$ (Fig. \ref{fig:schematic}B). The position vector in this case is given by

\begin{equation}
\textbf{r}(s,\theta)=r(s)\textbf{e}_{r}(\theta)+z(s) \textbf{k}.
  \label{eq:position vector}
\end{equation}

where ($\mathbf{e}_{r},\mathbf{e}_{\theta},\mathbf{k}$) form an orthogonal coordinate basis and $r(s)$ and $z(s)$ are the radius and elevation from the axis of revolution and base plane respectively. The complete parametrization is given in the SOM.  Since $(r')^2+(z')^2=1$, where $(')$ denotes derivative with respect to the arc length, we define an angle $\psi$ made by the tangent along the arc length with the horizontal such that $r'(s)=\cos\psi$ and $z'(s)=\sin\psi$.
The traction acting on a curve of constant $z$ is given by

\begin{equation}
\tilde{\textbf{f}}=\tilde{f}_{\nu} \bm{\nu}+\tilde{f}_n \textbf{n}.
\label{eq:traction_decomposition}
\end{equation}

\noindent where $\tilde{f}_n = (\tau W_{K})^{'} - (\frac{1}{2}W_{H})_{,\nu} - (W_{K})_{,\beta} \tilde{b}^{\alpha \beta} \nu_{\alpha}$  and $ \tilde{f_{\nu}} = W + \lambda - \kappa_{\nu} M$. Here, $\tilde{f}_{n}$ and $\tilde{f}_{\nu}$  represent the tractions normal and tangential to the membrane and are the curvature gradient energy per unit length and surface energy per unit length respectively \cite{rangamani2013interaction,agrawal2009boundary}.


Using $W$ as given in Eq. (\ref{eq:Helfrich})  and simplifying $\tilde{f}_n$ and $\tilde{f}_{\nu}$ equations, we obtain (see SOM for full derivation)

\begin{subequations}
\begin{align}
&\tilde{f}_n=-\kappa(H'-C'),\label{eq:traction-normal}\\ &\tilde{f}_{\nu}=\kappa(H-C)(H-C-\psi')+\lambda.
\label{eq:traction-tangential}
\end{align}
\end{subequations}




These equations, Eqs. (\ref{eq:traction-normal},\ref{eq:traction-tangential}), are our first result. We first provide a physical interpretation for these quantities, which have units of force per unit length or, equivalently, energy per unit area. First, the normal traction, $\tilde{f_n}$, represents the membrane response to deviations in the curvature gradient from the gradient in spontaneous curvature. A negative traction means that the membrane's reaction is in the opposite direction of the applied force. Second, the tangential traction, $\tilde{f}_{\nu}$, encompasses the local membrane tension as well as the deviation of the mean curvature from the spontaneous curvature. A positive traction indicates a tensile stress while a negative traction indicates a compressive stress. And third, as a sanity check for the model, in the absence of bending rigidity, we recover the tractions acting on the edge of a liquid droplet, with contribution from the membrane tension term only \cite{israelachvili2015intermolecular}. We now apply Eqs. (\ref{eq:traction-normal}, \ref{eq:traction-tangential}) to the formation of membrane tubes and buds.

\section*{Results}

\begin{figure}[t!]
\centering
\includegraphics[width =5in]{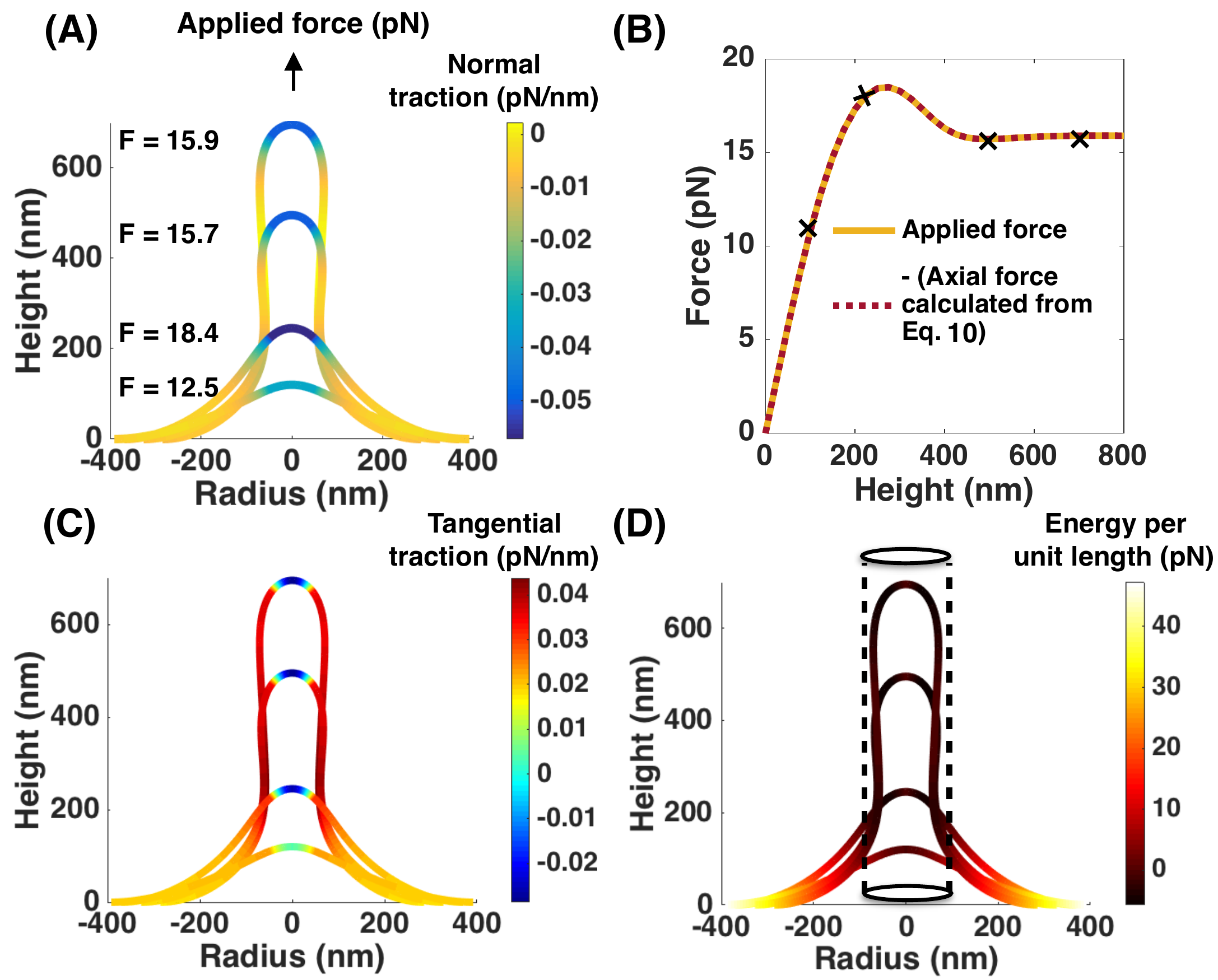}
\caption{Characterization of membrane tether using traction distributions. Here, we apply a point load at the pole and simulate the pulling of a membrane tether with membrane tension of $0.02 \,\mathrm{pN/nm}$, and bending modulus of $320$ $\mathrm{pN \cdot nm}$. (A) Normal traction along the membrane for four different values of externally applied force. Errors at pole due to the L = 0 boundary condition are removed by using asymptotic expansions. 
The negative value of traction represents the membrane's response to the positive external force. (B) Magnitude of axial force (Eq. \ref{eq:force-axial}) calculated at the base plotted alongside external force vs. height of tether. An exact match between the applied force and the model calculation is observed.  (C) Tangential traction distributions along the four membrane shapes. The large positive traction (acting downward) along the cylindrical region represents resistance to membrane deformation as the tube is pulled out. The negative value (acting leftward) at the tube tip is due to change in sign of surface tension (Eq. \ref{eq:lambda}). (D) Energy per unit length Eq.  (\ref{eq:EPUL}) along the four membrane shapes shown in (A). Cylinder describes the equilibrium geometry calculated using $\mathrm{R_{0}}$ = $\mathrm{\frac{1}{2}\sqrt{\frac{\kappa}{ \lambda}}}$. We observe negative energy per unit length inside the cylinder and positive values on the outside. Here, the energy per unit length predicts an effective line tension of $\sim$ 3 $\mathrm{pN}$ at the neck. }
\centering
\label{fig:tubes_simulation}
\end{figure}

\subsection*{Formation of membrane tethers requires both normal and tangential tractions} 

Tether formation is fundamental to cellular processes such as development of the growth cone, endocytosis etc. and captures how cytoskeletal forces deform the membrane actively \cite{hochmuth1996deformation,dai1995mechanical,dai1999membrane}.  The formation of membrane tethers in response to a point load is a classic example of force-mediated membrane deformation \cite{smith2004pulling,roux2002minimal} and has been studied in many experimental \cite{raucher1999characteristics,waugh1982surface,heinrich1999vesicle} and theoretical systems \cite{derenyi2002formation,walani2015endocytic,powers2002fluid}. In order to validate the expression for tractions (Eqs. \ref{eq:traction-normal}, \ref{eq:traction-tangential}) and to identify how normal and tangential tractions contribute to the formation of tethers, we conducted simulations mimicking the application of a local force at the pole. We used the following procedure - first, we obtained equilibrium membrane shapes for different values of a point load from simulations. Second, we used the simulated shapes and ran them through an image analysis algorithm - recalculating geometric parameters such as mean curvature, curvature gradient, and angle $\psi$. Finally, we used these geometric parameters to calculate the traction distribution (Eqs. \ref{eq:traction-normal}, \ref{eq:traction-tangential}) at every point, for predetermined values of surface tension. To resolve the calculation of tractions at the pole, we derived an asymptotic expression that allowed us to approximate the normal and tangential tractions for small arc length (see SOM for details). 

In Fig. \ref{fig:tubes_simulation}A, we plot the normal traction distribution along four equilibrium shapes generated by point force values specified next to each profile. The measured values highlight a distinct region along the tether cap with a large negative value, illustrating large curvature gradients there. The membrane curves away from the applied force along the region over which it is applied and conforms to a stable cylindrical geometry along the rest of the tether and a flat region at the base. Likewise, the tangential traction distribution features a large positive value along the cylindrical portion of the tether (Fig. \ref{fig:tubes_simulation}C) - the membrane resists stretching as the tube is pulled out. The tether cap shows negative values due to negative values of surface tension over the region of applied force. We can compare these tractions to dynamic boundary conditions at a fluid interface in the normal and tangential direction. 

The normal balance is an effective force balance between pressure and surface tension that assumes the form of the Young-Laplace equation  while the tangential balance equates to the gradient of surface tension \cite{Batchelor:1967ay}. Our expressions for tractions (Eqs. \ref{eq:traction-normal}, \ref{eq:traction-tangential}) reduce to their corresponding fluid analogues for negligible membrane rigidity and pressure difference. We can then interpret the normal and tangential traction as follows -- the tangential traction distribution tracks the gradient in `effective' surface tension while the normal traction distribution contains information regarding a force balance performed normal to the membrane at every point. 

We use this information to calculate membrane forces generated in response to the axial point load applied in our simulation. To do this, we write a general force balance in the presence of externally applied forces as

\begin{equation}
\underbrace{\int_{\omega} p\textbf{n} da}_{\text{\shortstack{Force due to pressure\\  acting on the surface}}}  {+} \underbrace{\int_{\partial \omega}\tilde{\textbf{f} }dt}_{\text{\shortstack{Traction force\\  along the boundary}}} {+} \underbrace{\textbf{F}}_{\text{External force} }=0.
\label{eq:force balance2}
\end{equation}

To find the axial force across the membrane, we first rewrite Eq.(\ref{eq:traction_decomposition}) in terms of $\textbf{e}_{r}(\theta)$ and $\textbf{k}$ as
 
\begin{equation}
\tilde{\textbf{f}}=\underbrace{(\tilde{f}_\nu\cos\psi-\tilde{f}_n\sin\psi)}_{\text{Radial traction}}\textbf{e}_{r}(\theta)+\underbrace{(\tilde{f}_\nu\sin\psi+\tilde{f}_n\cos\psi)}_{\text{Axial traction}}\textbf{k}.
\label{eg:traction1}
\end{equation}



The axial and radial components (Eqs. \ref{eq:s-traction-axial}, \ref{eq:s-traction-radial}) plotted along the equilibrium shapes are shown in Fig. \ref{Derenyi_axial_radial}. We then integrate the axial component of Eq. (\ref{eg:traction1}) along the circumference of the bounding curve $\partial \omega$ to obtain 

\begin{equation}
\tilde{F}_z=2 \pi r \Big[ \underbrace{\kappa H'\cos\psi+\kappa H(H-\psi')\sin\psi}_{\text{Bending contribution}}
+\underbrace{\lambda\sin\psi}_{\text{\shortstack{Tension\\contribution }}} \Big] 
\label{eq:force-axial},
\end{equation}

\noindent where $\tilde{F}_{z}$ is the axial force generated in response to the external load. 

We find that the negative of Eq. (\ref{eq:force-axial}) evaluated at the base of the geometry exactly traces the curve drawn by the external force (Fig. \ref{fig:tubes_simulation}B). Thus, the traction distributions act as an intermediary step between shape transitions and are sufficient to compute externally applied axial forces. This force match can be recreated for simulations with pressure by modifying our expression for force (Eq. \ref{eq:s-axial Force-variable moduli}, see Figs. \ref{tubewithpress_nolam}, \ref{tubewithpress}). Likewise, axial traction can be matched to a combination of traction due to pressure and external force (Eq. \ref{eq:s-axial Force in traction form}), shown in Fig. \ref{tractionmatch} for a tether pulling simulation against a large pressure of 1 MPa.    

In tether formation, besides the emergent axial traction in response to an applied force, we found that radial stresses play an important role in squeezing the membrane neck and holding the cylindrical configuration during membrane elongation. The energy per unit length, $\xi$, associated with this circular deformation can be found by integrating the radial traction  in Eq. (\ref{eg:traction1}) along the curve $\partial \omega$ (Fig. \ref{fig:schematic}B). This gives

\begin{equation}
\xi=2\pi r \Big[\underbrace{\kappa H(H-\psi')\cos\psi}_{\text{\shortstack{Curvature\\contribution }}}+\underbrace{\lambda \cos\psi}_{\text{\shortstack{Tension\\contribution }}}+\underbrace{\kappa H'\sin\psi}_{\text{\shortstack{Curvature\\ gradient\\contribution }}}\Big ].
\label{eq:EPUL}
\end{equation}

$\xi$ can be interpreted as an `effective' line tension, shown in Fig. \ref{fig:tubes_simulation}D. Whereas line tension computes the force acting at the boundary of two interfaces - e.g. inward force for a liquid droplet on a hydrophobic substrate and an outward force on a hydrophilic substrate \cite{buehrle2002impact} - the `effective' line tension predicts the general force acting at every point along the membrane shape, regardless of a phase boundary. Consequently, the point of zero line tension calculates the equilibrium geometry, shown as the dotted cylinder - the radius of which can also be evaluated by minimizing the free energy, giving $\mathrm{R_{0}}$ = $\frac{1}{2}\mathrm{\sqrt{\frac{\kappa}{ \lambda}}}$ \cite{derenyi2002formation}. Here, the equilibrium cylinder has no curvature gradient, leading to zero `effective' line tension. Measured values of energy per unit length inside the cylinder are negative while those outside are positive, indicating that the `effective' line tension determines the extent of deviation from the equilibrium geometry. Additionally, the value of $\xi$ at the neck is $\sim$ $ 3 \mathrm{pN}$, providing an estimate of the effective line tension required to form a neck in tethers.

\begin{figure}[t!]
\centering
\includegraphics[width=5in]{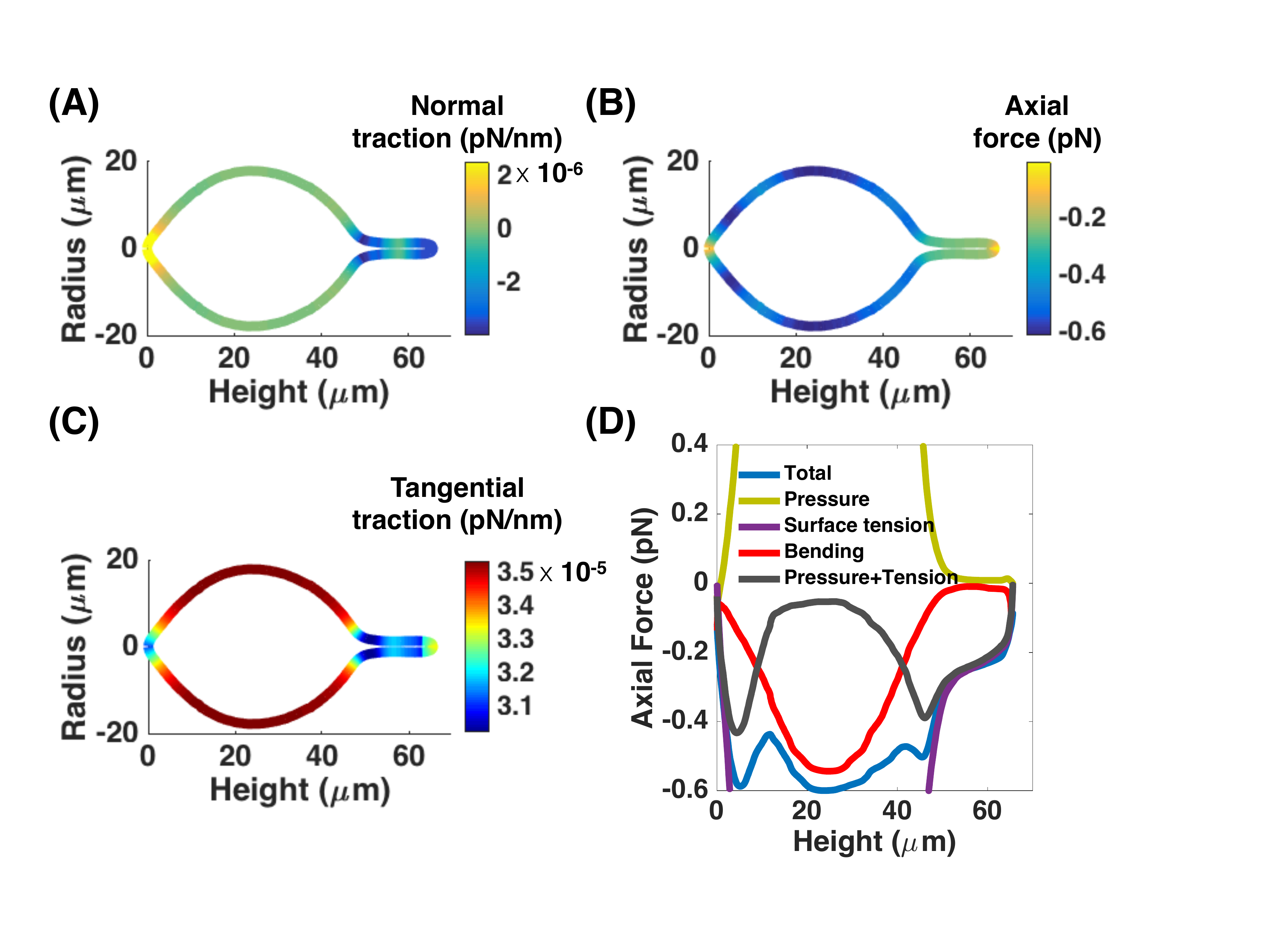}
\caption{Tractions calculated along vesicle shapes can predict external forces applied in experiments. We analyzed previously published  \cite{lee2008membrane} images of vesicles in force-extension experiments and calculated normal traction, tangential traction  Eq. (\ref{eq:traction-normal}, \ref{eq:traction-tangential})  and axial force (Eq.  \ref{eq:force-axial}). The parameters used were pressure p = $3.4 \,\mathrm{mPa}$, spontaneous curvature $C = -0.27 \,\mathrm{(\mu m)^{-1}}$, bending  modulus $\kappa = 0.85 \,\times 10^{-19} \,\mathrm{J}$ and surface tension $\lambda = 7.4 \,\mathrm{k_{B}T/(\mu m)^{2})}$ \cite{lee2008membrane}. (A) and (C) Normal and tangential traction distributions along a vesicle of height 70 $\mu$ m.  Normal traction is large and negative at the pole while tangential traction is offset by a large surface tension and is large and positive along cylindrical part of the tether. The rest of the shape shows negligible normal traction and constant tangential traction, because the of the near-spherical geometry, with deviations close to the base. These deviations can be attributed to the anchoring of the vesicle to a stationary bead that alters the vesicle shape at the pole. (B) Axial force along same vesicle. We observe axial force of $\sim -0.55 \,\mathrm{pN}$ along most of the vesicle and a smaller axial force of $\sim -0.25 \,\mathrm{pN}$ at the tether. (D) Axial force and its components plotted vs height of vesicle. The sub $\mathrm{pN}$ axial force obtained from our calculations matches experimentally observed values \cite{lee2008membrane}. Forces due to pressure and surface tension balance each other and combine with bending forces to give a near-constant value in the spherical regions.}
\centering
\label{fig:wiggins}
\end{figure}

\subsection*{Analysis of vesicle shapes gives insight into the distribution of normal and tangential tractions}
We next asked if the shape of a vesicle obtained from existing data could provide information on the normal and tangential tractions. We used previously published images of vesicles with tethers \cite{lee2008membrane} to calculate the traction distributions. Briefly, the grayscale images were imported into MATLAB and the outline of the vesicle was traced. Then the geometric parameters were calculated as before to obtain the distribution of normal and tangential tractions along the imported geometry shown in Fig. (\ref{fig:wiggins}A,C). We observe that the distribution of tractions along the tether resembles that of Fig. \ref{fig:tubes_simulation} -- normal traction is large and negative along the tether cap and negligible along the cylindrical portion while the tangential traction is positive along the tether and resists membrane stretch. The differences begin from the base of the tether as the membrane conforms to a vesicle geometry. Here, we see that normal traction becomes negligible and tangential traction assumes a constant value, indicating a stable spherical geometry. However, the base of the vesicle shows large normal traction and smaller tangential traction. This represents the response to the shape that the membrane is forced to take by an optically trapped bead anchoring the GUV at its base. To calculate axial force along the geometry, Eq. (\ref{eq:force-axial}) was modified for pressure difference (Eq. \ref{eq:s-axial Force}) and implemented at every point along the curve. In Lee et al. \cite{lee2008membrane}, the applied force was reported as $\sim 0.6 \mathrm{pN}$ along the entire membrane. Here, we found that the axial force matched closely those reported values along the curved portion with an average force of $0.55 \,\mathrm{pN}$, and the force along the tether was $0.25 \mathrm{pN}$ (Fig. \ref{fig:wiggins}B).  Furthermore, the different contributions of pressure, tension, and bending follow the same profile as that reported in \cite{lee2008membrane} (Fig. \ref{fig:wiggins}D), indicating that we can not only extract the applied forces on the membrane using shape information but also evaluate contributing terms.

 \begin{figure}[t!]
\centering
\includegraphics[width=5in]{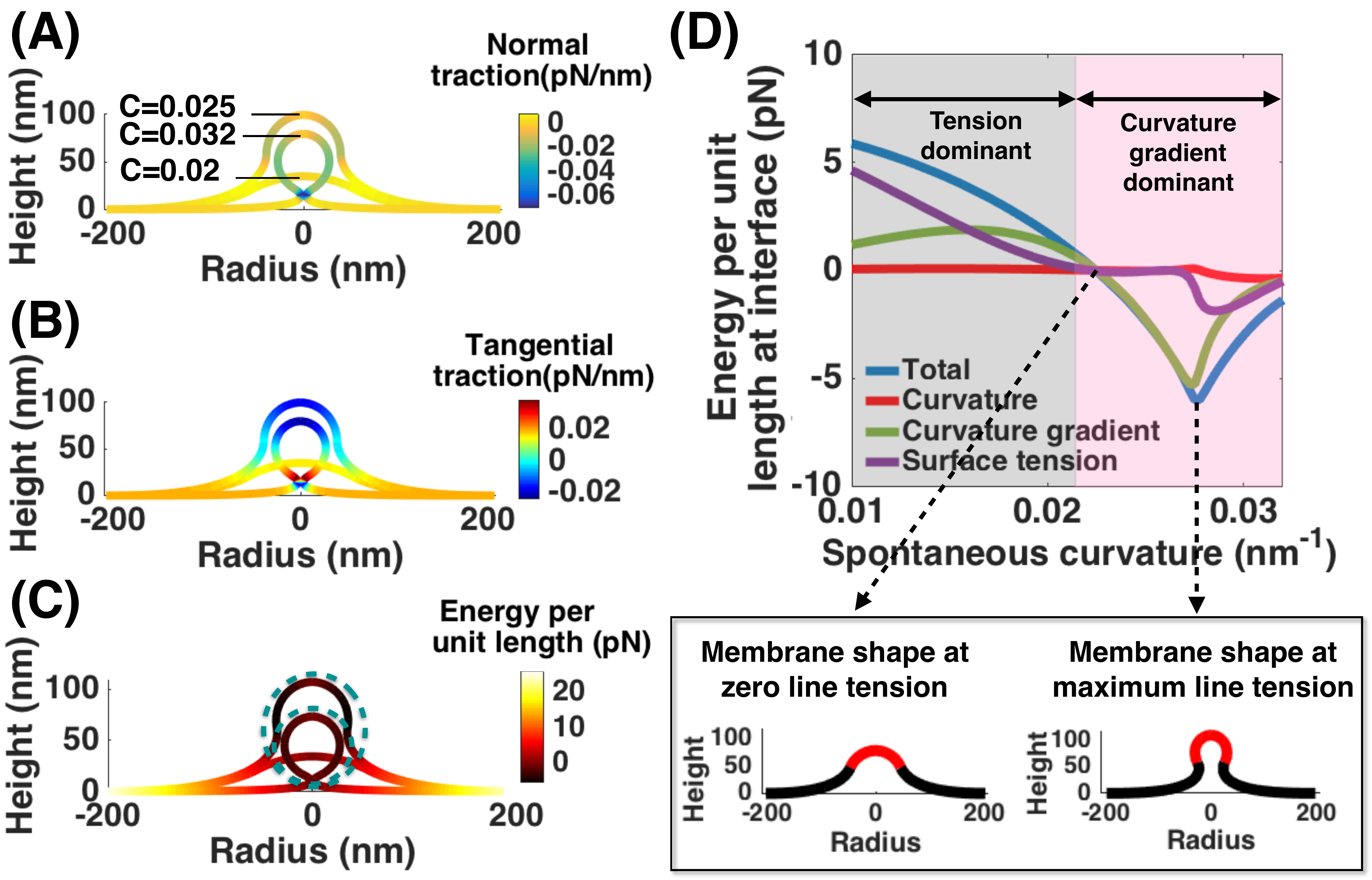}
\caption{Characterization of bud formation in heterogeneous membranes using traction distributions. In these simulations, a constant area of spontaneous curvature $\mathrm{A}=$10,053 $\mathrm{nm^2}$ is developed from the center of an initially flat patch while the magnitude of spontaneous curvature increases from $C=0$ to $C=0.032 \,\mathrm{nm^{-1}}$. Membrane tension is set to $0.02 \,\mathrm{pN/nm}$ at the edge, the bending modulus is constant at $320 \,\mathrm{pN \cdot nm}$ and the pressure difference across the membrane is assumed to be zero \cite{hassinger2017design}. (A) Negative normal traction along the curved bud represents the membrane tendency to form a narrow neck in the area of disorder lipids. (B) Tangential traction changes sign from positive in shallow bump to negative along the cap  once the bud takes on a  U-shape. The discontinuity at the neck is mostly due to surface tension heterogeneity at the edge of the region of spontaneous curvature. (C) Energy per unit length distribution for three different shapes. The dashed line circles characterize the equilibrium vesicle shapes at $C=0.032 \,\mathrm{nm^{-1}}$ (smaller circle) and $C=0.025 \,\mathrm{nm^{-1}}$ (larger circle). Negative energy per unit length is only observed inside the equilibrium shapes. (D) Variation of energy per unit length and its components at the interface with change in spontaneous curvature. Two regimes are observed: (1) surface tension dominated regime for small values of spontaneous curvature where energy per unit length is positive, (2) curvature gradient dominated regime for large vales of spontaneous curvature where energy per unit length is negative - necessary to form narrow necks. The membrane configurations are shown for two spontaneous curvature magnitudes  $C=0.02 \,\mathrm{nm^{-1}}$, where energy per unit length at interface is zero and $C=0.025 \,\mathrm{nm^{-1}}$, where energy per unit length is maximum. The red domains show the region of spontaneous curvature for the corresponding shapes.}
\centering
\label{fig:bud_simulation}
\end{figure}

\subsection*{Formation of buds due to spontaneous curvature is regulated by emergent line tension}
Phase separation and lipid domains are classical mechanisms of bud formation and vesiculation \cite{richmond2011forming,hassinger2017design}. Previously, we and others have shown that the heterogeneity on the membrane can be modeled using a spontaneous curvature field \cite{rangamani2014protein,agrawal2009modeling,steigmann1999fluid}. We used this model (details in the SOM) to investigate the nature of membrane tractions generated by spontaneous curvature in bud formation. In the first step, we set up simulations for a constant area of spontaneous curvature field $\mathrm{A}=10,000 \mathrm{nm^2}$ developed from the center of an initially flat patch. The induced spontaneous curvature due to asymmetry was increased from $C=0$ $\mathrm{nm^{-1}}$ to $C=0.032$ $\mathrm{nm^{-1}}$ and the region of phase separation was modeled using a hyperbolic tangent function. We then chose three distinct shapes -- a shallow bump,  a U shaped bud, and a closed bud as inputs for our image analysis algorithm. In Fig. \ref{fig:bud_simulation}A, we plot normal traction distribution along these three shapes. Negative normal traction along the area of the spontaneous curvature is an indicator of  sharper change in mean curvature compared to the applied asymmetry. This discrepancy becomes larger with increasing magnitude of spontaneous curvature. At the neck, where $\psi=\frac{\pi}{2}$, normal traction is maximum and acts purely inward, representing the tendency of the membrane to form small necks.  

The sudden fall from positive to negative tangential traction at the necks highlights the critical role of the gradient in tangential traction in formation of narrow necks and instability \cite{walani2015endocytic,hassinger2017design} (Fig. \ref{fig:bud_simulation}B). For tent-like small deformations, the tangential traction is positive throughout indicating that the membrane resists bending deformation. However, in the U shaped and closed buds, the negative tangential traction along the cap acts to pull the membrane and favors adopting a highly curved shape.

In the previous section we showed that a tether can be formed by applying a point load. Here, we asked whether it is possible to replace the heterogenity with a load and still form a bud. To answer this question we used a modified version of Eq. (\ref{eq:force-axial}) with spontaneous curvature (\ref{eq:s-axial Force}) and calculated the axial force at the base. For all three shapes, the axial force was negligible ($\sim 10^{-4}$ $\mathrm{pN}$) indicating that bud formation is solely controlled by radial tractions (see Fig. \ref{fig:bud-axial-radial}). The energy per unit length associated with this deformation can be evaluated by a modified version of Eq. (\ref{eq:EPUL}) including spontaneous curvature (Eg. \ref{eq:s-EPUL}) (Fig. \ref{fig:bud_simulation}C). The dashed circles represent the equilibrium spherical vesicles calculated by Helfrich energy minimization ($R_0=\frac{\kappa C}{\lambda+\kappa C^2}$) \cite{hassinger2017design}. Each equilibrium vesicle divides the space into two domains; (\textit{i}) the membrane inside the vesicle with negative energy per length that bends to form a bud (\textit{ii}) the membrane outside the vesicle with positive surface tension that resists deformation.


Previously, both modeling and experimental studies have shown that line tension in heterogeneous membranes can be sufficient for scission in endocytosis \cite{liu2006endocytic} and the formation of buds in vesicle experiments \cite{baumgart2003imaging,baumgart2005membrane}. We use a modified expression for energy per unit length (Eq. \ref{eq:s-EPUL}) to estimate line tension at the interface of the two domains. Through the process of bud formation, line tension undergoes a sign change from positive (acting outward) to negative (acting inward), effectively transitioning from a tension-dominated regime to a curvature gradient-dominated regime while the term due to curvature is almost zero (Fig. \ref{fig:bud_simulation}D). This transition from positive to negative line tension with increasing value of spontaneous curvature is also observed in trans-membrane proteins \cite{dan1998effect}. Energy per unit length at the interface varies between -5 $\mathrm{pN}$ to 5 $\mathrm{pN}$, which is the reported order of interfacial line tension between coexisting phases in lipid bilayers \cite{liu2006endocytic,lipowsky1992budding}. We also see that once the overhang develops, closing the neck requires a smaller line tension. In this simulation, we set surface tension at the boundary to be $\lambda_0=0.02$ $\mathrm{pN/nm}$. However, this value can vary based on the type of reservoirs affecting the line tension at the interface (see Fig. \ref{fig:tensioncomp}).

\begin{figure}
\centering
\includegraphics[width=5in]{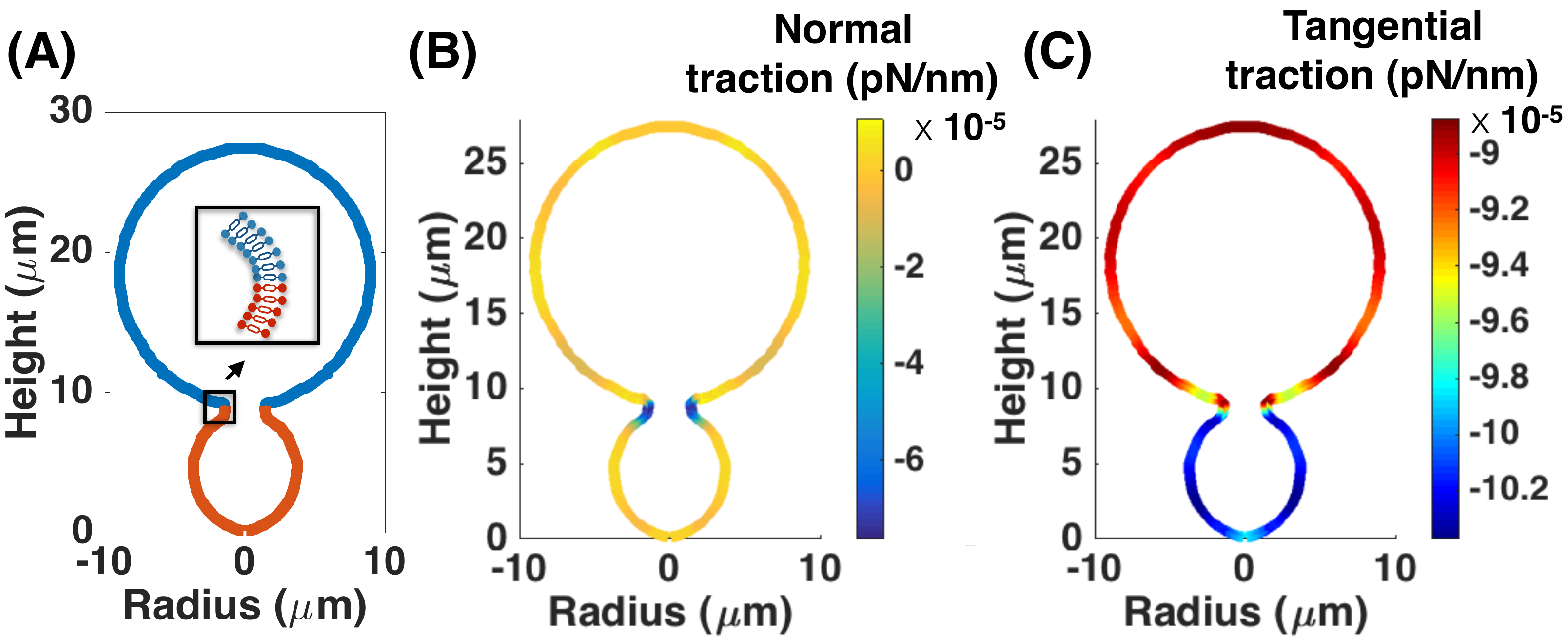}
\caption{Tractions can predict experimentally observed values of line tension at an interface. We used published two-photon microscopy images of axially symmetric vesicles with fluid phase coexistence and calculated the normal and tangential tractions. The constant parameters used were   p = $2.8 \,\times 10^{-2} \,\mathrm{N/m^2}$, surface tension of disordered phase $\lambda_d = -1.03 \,\times 10^{-4} \,\mathrm{mN/m}$, surface tension of ordered phase  $\lambda_o = -0.91 \,\times 10^{-4} \,\mathrm{mN/m}$, bending modulus of disordered phase $\kappa_d = 10^{-19} \,\mathrm{J}$, ratio of bending modulii $\kappa_o/\kappa_d = 5$ and absolute difference in gaussian moduli $\Delta \kappa_G = 3.6 \times 10^{-19} \,\mathrm{J}$ \cite{baumgart2005membrane}. Spontaneous curvature was estimated by taking inverse of radius of the largest sphere that fits inside each region. (A) Geometry of a vesicle with lipid ordered phase $L_{o}$ (Blue)  and lipid disordered phase $L_{d}$ (Red). (B) Normal traction distribution shows a large negative normal traction at the interface. Computing the energy per unit length at this point where $\psi=\frac{\pi}{2}$ and $\tilde{f}_n=\tilde{f}_r$) gives a value of 0.7 $\mathrm{pN}$, which is in order of the experimentally observed value of line tension 0.67 $\mathrm{pN}$ \cite{baumgart2005membrane}. (C) Tangential traction along the vesicle shows negative value everywhere. $L_{d}$ phase has a larger magnitude of tangential traction while $L_o$ phase has a smaller value, leading to a gradient at the interface.}
\centering
\label{fig:baumgart}
\end{figure}

\subsection*{Analysis of experimental shapes can predict empirical values of line tension}
We next asked if the expressions for energy per unit length (Eq. \ref{eq:EPUL}) could be extended to the estimation of line tension at the interface of two domains observed in previously published data. We used images from \cite{baumgart2005membrane} and calculated normal and tangential tractions along the vesicle shape following the same procedure as before. The constant parameters in Eqs. (\ref{eq:s-general-traction-normal}, \ref{eq:s-general-traction-tangential}) are chosen based on reported values in this paper, while the spontaneous curvature for each vesicle is estimated by the inverse of radius of the largest circle fitted inside each region. A rendering of the original image is shown in Fig. \ref{fig:baumgart}A. The normal traction distribution along the membrane showed that it is almost zero everywhere except at the neck, which is the interface of the lipid disordered and ordered phase (Fig. \ref{fig:baumgart}B). Calculating the corresponding energy per unit length at the neck (Eq. \ref{eq:s-EPUL-variable moduli})  gives a value of $0.7 \mathrm{pN}$, comparable to the experimentally obtained value of $0.67 \mathrm{pN}$ \cite{baumgart2005membrane}. Thus, the line tension that is acting at the interface of two lipid domains or between a protein-coated membrane and a bare membrane can be estimated from the information contained in the shape of the vesicle and a few material parameters. Furthermore, the tangential traction distribution is characterized by a gradient at the neck in the region of fluid phase coexistence, a key factor that has not been emphasized in previous studies. Fig. \ref{Baumgart A} shows similar behavior for two other experimental vesicle shapes \cite{baumgart2005membrane}.  

\section*{Discussion}

In this study, we have presented general formulae for the calculation of normal and tangential tractions along a rotationally symmetric membrane shape. As a test of these formulae, we verified their ability to extract the force-displacement profile from simulations of membrane tubulation. We then applied our formula to previously published experiments of tether pulling from vesicles and obtained strong agreement with the axial force distribution calculated through an alternate method \cite{lee2008membrane}. 
We also used these formulae to demonstrate that spontaneous-curvature-mediated budding is driven by both normal and tangential tractions. 
We further show that, in the case of budding in phase-separated GUVs \cite{baumgart2005membrane}, energy per unit length is greatest at the interface between two phases, indicating that line tension can be understood to be a special case of the general phenomenon of radial traction-mediated budding.

Importantly, we have demonstrated that the normal and tangential tractions on the membrane are emergent properties that arise as a consequence of different mechanisms of curvature generation. Moving forward, these formulae provide a general tool for the analysis of forces acting on membranes whether that be in reconstituted systems or in cells. In particular, we expect that analysis of the force distributions along membrane shapes will provide insight into how various input mechanisms are transduced into forces on the membrane that shape membrane curvature (Fig. \ref{fig:schematic}).

An observation that we have made here is that there is a fundamental difference in the axial and radial tractions between membrane deformations generated by an axially applied force (e.g. tethers) as compared to isotropic spontaneous curvature (e.g. buds). One particular example where these two modes of curvature generation come together is in clathrin-mediated endocytosis where yeast have tubular-like invaginations but mammalian cells have spherical pits \cite{conibear2010converging,idrissi2008distinct}. This suggests that the shape-traction relationship is unique.

At present it remains very difficult to dissect the combination of molecular mechanisms that are responsible for the shape of curved cellular membrane structures such as membrane buds, sheets, tubules, and filopodia \cite{baumgart2011thermodynamics}. For example, during the formation of coated vesicles, which are among the best studied curved cellular structures, diverse mechanisms including protein scaffolding, hydrophobic insertion, and protein crowding have each been implicated \cite{stachowiak2013cost}. In contrast to the analysis of cellular structures, in vitro studies on purified proteins and lipids allow us to examine the impact of individual proteins and mechanisms on membrane shape. These data, which are abundant in the literature \cite{dannhauser2012reconstitution,frost2009bar,snead2017membrane}, can be compared to the model proposed here and used to refine it. Once refined, the model can then be used to analyze the shapes of complex structures in cells, which likely include contributions from multiple mechanisms.

Recently electron microscopy data has advanced considerably in resolution, enabling the visualization of membranes shapes in cells in unprecedented detail \cite{avinoam2015endocytic,mccullough2015structure}. Comparing the model proposed here to these data represents a new opportunity to dissect the physical mechanisms that come together to create the diversity of membrane shapes observed in cells. Additionally, the sensitivity condition in our model requires only that the shapes captured be continuous. Beyond this, we expect that the approach will be powerful for understanding how cells regulate their function through geometry, mechanics, and signaling \cite{rangamani2013decoding}.


Ultimately, we believe that understanding the normal and tangential tractions on the membrane and their projection in the axial and radial directions will provide a framework for the understanding and rational design of novel mechanisms for membrane bending. For example, it has been demonstrated that PEGylation of lipids \cite{lee2011coarse}, amphiphilic block copolymers \cite{lim2017spontaneous}, and protein crowding \cite{snead2017membrane} can curve and even induce scission of artificial lipid bilayers. Additional terms may be necessary in our formulae to account for these novel modes of membrane bending, but fundamentally there is no change to the principle of force balance that allows us to use the membrane shape as a readout of the forces. Thus, studying this inverse problem can lead to an understanding of the integration of curvature-generating mechanisms into applied forces on the membrane. Furthermore, this approach allows us to connect synthetic chemistry to mechanochemistry in order to engineer tools that can manipulate membrane curvature.

\textbf{Acknowledgment} work was supported by  ARO W911NF-16-1-0411, AFOSR FA9550-15-1-0124, and NSF PHY-1505017  grants to P.R. J.C.S. was supported by NIH R01GM112065. R.V. was supported by the UCSD Frontiers of Innovation Scholars Program (FISP) G3020. H.A. was supported by a fellowship from the Virtual Cell Consortium, a program between UCSD and the Scripps Research Institute. J.E.H. was supported by the Department of Defense (DoD) through the National Defense Science \& Engineering Graduate Fellowship (NDSEG) Program. The authors would also like to thank Prof. George Oster and Prof. David Steigmann for initial discussions.



\bibliography{traction} 
\bibliographystyle{plain} 

\newpage

\setcounter{equation}{0}
\setcounter{section}{0}
\setcounter{figure}{0}
\setcounter{table}{0}
\setcounter{page}{1}
\renewcommand{\theequation}{S\arabic{equation}}
\renewcommand{\thefigure}{S\arabic{figure}}

\title{ {\bf \hf
Supplementary material \\ \bf \hf `The role of traction in membrane curvature generation'}}

\author{\bf \hf Haleh Alimohamadi$^1$, Ritvik Vasan$^1$, Julian E. Hassinger$^2$\\ \bf \hf Jeanne C. Stachowiak$^3$, Padmini Rangamani$^{1*}$}
\date{\small \hf
$^1$Department of Mechanical and Aerospace Engineering, \\
University of California San Diego, La Jolla CA 92093\\
$^2$Biophysics Graduate Program, University of California, Berkeley, CA 94720\\
$^3$Department of Biomedical Engineering, University of Texas at Austin, Austin, TX 78712\\
\medskip 
$^*$To whom correspondence should be addressed. e-mail: prangamani@ucsd.edu \\
\medskip 
\today}

\maketitle
\tableofcontents


\newpage

\section{Model Development}\label{sec:model-development}
\subsection{Assumptions}\label{sec:assumption}

\begin{itemize}
  \item  Membrane curvature generated due to forces or protein-induced spontaneous curvature is much larger than the thickness of the bilayer. Based on this assumption, we model the lipid bilayer as a thin elastic shell with a bending energy given by the Helfrich-Canham energy, which is valid for radii of curvatures much larger than the thickness of the bilayer \cite{helfrich1973elastic,miller2015calm}.
 
\item For current purposes, we neglect the surrounding fluid flow or inertial dynamics and assume that the membrane is at mechanical equilibrium at all times \cite{steigmann2003variational}. This assumption is commonly used in the modeling of membrane curvature to keep the mathematics tractable \cite{steigmann1999fluid}.

\item The membrane is incompressible because the energetic cost of stretching the membrane is high \cite{rawicz2000effect}. This constraint is implemented using a Lagrange multiplier \cite{rangamani2013interaction,rangamani2014protein} as discussed in Section (\ref{sec:equilibrium}). 


\item Finally, for simplicity in the numerical simulations, we assume that the membrane in the region of interest is rotationally symmetric (Fig. \ref{fig:schematic}). 
\end{itemize}

\subsection{Equilibrium equations}\label{sec:equilibrium}

Force balance on the membrane can be written as

\begin{align} 
\nabla \cdot \bm{\sigma}+p\textbf{n}=\textbf{f}, 
\label{eq:s-Newton}
\end{align}

\noindent where $\nabla \cdot$ is surface divergence, $\bm{\sigma}$ is the stress vector, p is the pressure difference between the inside and outside of the volume bounded by membrane, and $\textbf{f}$ is any externally applied force per unit area on the membrane. By introducing the covariant derivative as $()_{;\alpha}$, the surface divergence in Eq. (\ref{eq:s-Newton}) can be rewritten as \citep{steigmann1999fluid}

 \begin{align} 
\nabla \cdot \bm{\sigma}=\bm{\sigma}^{\alpha}_{;\alpha}=(\sqrt{a})^{-1} (\sqrt{a}\bm{\sigma}^{\alpha})_{,\alpha},
\label{eq:s-divergence}
\end{align}

\noindent where $a$ is determinant of the first fundamental form metric $a_{\alpha \beta}$. The surface stresses in Eq. (\ref{eq:s-Newton}) can be split into normal and tangential component given by

 \begin{align} 
\sigma^{\alpha}= T^{\alpha}+S^{\alpha} \textbf{n},
\label{eq:s-stress decomposition}
\end{align}

\noindent where

\begin{align} 
\textbf{T}^{\alpha} =T^{\alpha \beta} \textbf{a}_{\beta}, \quad \quad T^{\alpha \beta}=\sigma^{\alpha \beta}+b^{\beta}_{\mu}M^{\mu \alpha}, \quad  \quad S^{\alpha}=-M^{\alpha \beta}_{; \beta}.
\label{eq:s-stress1}
\end{align}

The two tensors $\sigma^{\alpha \beta}$ and $M^{\alpha \beta}$ can be expressed by the derivative of $F$, the energy per unit mass, with respect to the coefficients of the first and second fundamental forms,  $a_{\alpha \beta}$, $b_{\alpha \beta}$, respectively \cite{rangamani2013interaction,steigmann1999fluid}

\begin{align} 
\sigma^{\alpha \beta}=\rho (\frac{\partial F(\rho,H,K;x^{\alpha})}{\partial a_{\alpha\beta}}+\frac{\partial F(\rho,H,K;x^{\alpha})}{\partial a_{\beta\alpha}}), \quad M^{\alpha\beta}=\frac{\rho}{2} (\frac{\partial F(\rho,H,K;x^{\alpha})}{\partial b_{\alpha\beta}}+\frac{\partial F(\rho,H,K;x^{\alpha})}{\partial b_{\beta\alpha}}),
\label{eq:s-F derivative}
\end{align}

\noindent where $\rho$ is the surface mass density. $H$ and $K$ are mean and Gaussian  curvatures  given by

\begin{align} 
H=\frac{1}{2} a^{\alpha \beta}b_{\alpha \beta}, \quad K=\frac{1}{2} \varepsilon^{\alpha \beta} \varepsilon ^{\lambda \mu} b_{\alpha \lambda}b_{\beta \mu}.
\label{eq:s-Curvatures}
\end{align}

Here $(a^{\alpha\beta})=(a_{\alpha \beta})$ is the dual metric and $\varepsilon ^{\alpha \beta}$ is the permutation tensor defined by $\varepsilon ^{12}=-\varepsilon ^{21}=\frac{1}{\sqrt{a}}, \varepsilon ^{11}=\varepsilon ^{22}=0$. 

Area incompressibility ($J=1$) constrain is implemented using a general form of free energy density per unit mass given as

\begin{align} 
F(\rho,H,K;x^{\alpha})=\tilde{F}(H,K;x^{\alpha})-\frac{\gamma (x^{\alpha},t)}{\rho}.
\label{eq:s-Lagrange multiplier}
\end{align}

Here $\gamma (x^{\alpha, t})$ is a Lagrange multiplier field required to impose invariance of $\rho$ on the whole of the surface (see \cite{steigmann1999fluid} for full derivation). Substituting $W=\rho \tilde{F}$ into Eq. (\ref{eq:s-Lagrange multiplier}) we get

\begin{align} 
\sigma^{\alpha \beta}=(\lambda +W)a^{\alpha \beta} -(2HW_H+2\kappa W_K)a^{\alpha \beta} +W_H\tilde{b}^{\alpha \beta}, \\
M^{\alpha \beta}=\frac{1}{2}W_H a^{\alpha \beta} +W_K \tilde{b}^{\alpha \beta}.
\label{eq:s-stress2}
\end{align}

\noindent where 

\begin{align} 
\lambda=-(\gamma+W).
\label{eq:s-lambdaa}
\end{align}

Combining Eqs. (\ref{eq:s-stress2}, \ref{eq:s-stress1}), and (\ref{eq:s-stress decomposition}) into Eq. (\ref{eq:s-Newton}) give the equations in normal and tangential equations as

\begin{align} 
p+\textbf{f} \cdot \textbf{n}=\Delta{\frac{1}{2} W_H}+(W_K)_{; \alpha \beta} \tilde{b}^{\alpha \beta} +W_H(2H^2- K) +2H(KW_K-W)-2 \lambda H.
\label{eq:s-normal}
\end{align}

and

\begin{align} 
N^{\beta \alpha}_{; \alpha}-S^{\alpha}b^{\beta}_{\alpha}=-(\gamma_{, \alpha} +W_K k_{, \alpha}+W_H H_{, \alpha}) a^{\beta \alpha}=(\frac{\partial{W}}{\partial{x^{\alpha}_{| exp}}}+\lambda_{, \alpha}) a^ {\beta \alpha} = \textbf{f}.\textbf{a}_s.
\label{eq:s-Tangential}
\end{align}
 
Here $\Delta (.)$ is the surface Laplacian and $()_{| exp}$ denotes the explicit derivative respect to coordinate $\theta^{\alpha}$

\subsubsection{Helfrich energy with constant bending and Gaussian moduli}

For a lipid bilayer with uniform bending and Gaussian moduli, we use a modified version of the Helfrich energy to account for the spatial variation of spontaneous curvature \citep{agrawal2009boundary,walani2015endocytic,rangamani2014protein},

\begin{align} 
W=\kappa (H-C(\theta^{\alpha}))^2+\kappa_G K\, 
\label{eq:s-Helfrich}
\end{align}

\noindent where W is the local energy density, C is the spontaneous curvature and $\kappa$ (bending modulus) and $\kappa_G$ (Gaussian modulus) are constant. It should be mentioned that Eq. (\ref{eq:s-Helfrich}) is different from the standard Helfrich energy by a factor of 2. We take this net effect into consideration by choosing the value of the bending modulus to be twice of the standard value of bending modulus typically used for lipid bilayers\cite{helfrich1973elastic}. Even though the membrane bending and Gaussian moduli need not necessarily be uniform due to composition variation along the membrane \cite{agrawal2011model,stachowiak2013cost}, assuming uniform $\kappa$ and $\kappa_G$ is  an acceptable simplification for classical simulations. In both the tube and bud simulations, we assumed that bending and Gaussian moduli are constants. The general form of equations with variable bending and Gaussian moduli is given in \ref{sec:variable}. 

At equilibrium, the integration of local energy density over the total membrane surface area $\omega$ gives the strain energy of the system written as

\begin{align} 
E=\int_{\omega}(\kappa (H-C(\theta^{\alpha}))^2+\kappa_G K)da, 
\label{eq:s-strainenergy}
\end{align}
 \noindent where E is total strain energy. Imposing area and volume conservation by Lagrange multipliers p and $\lambda$ gives
 
 \begin{align} 
E=\int_{\omega}\underbrace{(\kappa (H-C(\theta^{\alpha}))^2+\kappa_G K +\lambda) }_\text{Energy density}da-\underbrace{pV (\omega)}_\text{Pressure work}, 
\label{eq:s-energy density}
\end{align}

\noindent where V is the volume associated with the membrane surface. The strain energy in Eq. (\ref{eq:s-energy density}) can be split in three different components as

\begin{align} 
E=\underbrace{\int_{\omega}(\kappa (H-C(\theta^{\alpha})^2+\kappa_G K) da}_\text{Bending energy}+\underbrace{\int_{\omega} \lambda da}_\text{Surface tension work}-\underbrace{pV (\omega)}_\text{Pressure work}. 
\label{eq:s-energy density2}
\end{align}

Using the Helfrich energy function Eq. (\ref{eq:s-Helfrich}) in the balance of forces normal to the membrane (Eq. (\ref{eq:s-normal})) yields the ``shape equation,''
 
\begin{align} 
\kappa \underbrace{\Delta \left[\left(H - C \right)\right] + 2 \kappa \left( H - C \right) \left(2 H^2 - K \right) - 2 \kappa H \left( H - C \right)^2}_{\text{Elastic Effects}} = \underbrace{p + 2 \lambda H}_{\text{Capillary effects}} + \underbrace{\textbf{f} \cdot \textbf{n}}_{\text{Force due to actin}}, \label{eq:s-shape}
\end{align}


\noindent where $\lambda$ can be interpreted to be the membrane tension \cite{rangamani2014protein}. 

A consequence of heterogenous protein-induced spontaneous curvature, heterogeneous moduli, and externally applied force is that $\lambda$ is not homogeneous in the membrane \cite{agrawal2009boundary,hassinger2017design,rangamani2014protein}. Substituting Eq. (\ref{eq:s-Helfrich}) in the balance of forces tangent to the membrane Eq. (\ref{eq:s-Tangential}) gives the spatial variation of membrane tension,

\begin{align} 
\underbrace{\lambda_{, \alpha}}_{\text{\shortstack{Gradient of \\ surface tension}}} = \underbrace{2 \kappa \left( H - C \right) \frac{\partial{C}}{\partial{\theta^{\alpha}}}}_{\text{Protein-induced variation}} - \underbrace{\textbf{f} \cdot \textbf{a}_\alpha}_{\text{\shortstack{Force induced \\  variation}}}.
\label{eq:s-lambda}
\end{align}

\subsubsection{Helfrich energy with variable bending and Gaussian moduli} \label{sec:variable}

For a membrane with variable bending and Gaussian moduli the modified Helfrich energy in Eq. (\ref{eq:s-Helfrich}) becomes

\begin{align} 
W=\kappa (\theta^{\alpha}) (H-C(\theta^{\alpha}))^2+\kappa_G (\theta^{\alpha}) K\, 
\label{eq:s-variable-Helfrich}
\end{align}

\noindent where both moduli vary along the membrane. 
Substituting Eq. (\ref{eq:s-variable-Helfrich}) into Eqs. (\ref{eq:s-shape}) and (\ref{eq:s-lambda}) gives a complete form of the so-called `shape equation' and spatial variation of membrane tension given below

\begin{align} 
 \Delta \left[\kappa\left(H - C \right)\right] + 2H\Delta\kappa_G-(\kappa_G)_{;\alpha \beta}b^{\alpha \beta}+2 \kappa \left( H - C \right) \left(2 H^2 - K \right) - 2 \kappa H \left( H - C \right)^2 = p + 2 \lambda H + \textbf{f} \cdot \textbf{n}, 
 \label{eq:s-variable-shape}
\end{align} 

\begin{align} 
\lambda_{, \alpha} = 2 \kappa \left( H - C \right) \frac{\partial{C}}{\partial{\theta^{\alpha}}}-\frac{\partial \kappa}{\partial \theta^{\alpha}}(H-C)^2-\frac{\partial \kappa_G}{\partial \theta^{\alpha}}K- \textbf{f}\cdot\textbf{a}_\alpha.
\label{eq:s-variable-lambda}
\end{align}

Here, $b^{\alpha\beta}$ are components of the curvature tensor.


\subsection{Axisymmetric coordinates}
\subsubsection{Equation of motion for constant bending and Gaussian moduli}

We parametrize a surface of revolution (Fig. \ref{fig:schematic}) by

\begin{equation}
\textbf{r}(s, \theta) = r(s)\textbf{e}_{r}(\theta) + z(s)\textbf{k}.
\end{equation} 

We define $\psi$ as the angle made by the tangent with respect to the horizontal. This gives $r'(s) =  \cos(\psi)$, $z'(s) =  \sin(\psi)$, which satisfies the identity $(r')^2+ (z')^2 = 1$. Using this, we  define the normal to the surface as $\textbf{n}=-\sin\psi\textbf{e}_{r}(\theta)+\cos\psi\textbf{k}$, the tangent to the surface in the direction of increasing arc as $\bm{\nu}=\cos\psi\textbf{e}_{r}(\theta)+\sin\psi\textbf{k}$ and unit vector $\bm{\tau}=\textbf{e}_{\theta}$ tangent to the boundary $\partial \omega$ in the direction of the surface of revolution (see Fig. \ref{fig:schematic}).


This parametrization yields the following expressions for  tangential $(\kappa_{\nu})$ and transverse $(\kappa_{\tau})$ curvatures, and twist $(\tau)$:

\begin{equation}
\kappa_{\nu}=\psi^{'},\quad \kappa_{\tau}=r^{-1}\sin\psi,\quad \tau=0.
\label{eq:s-kappas}
\end{equation}

The mean curvature ($H$) and Gaussian curvature ($K$) are obtained by summation and multiplication of the tangential and transverse curvatures

\begin{equation}
H=\frac{1}{2}(\kappa_\nu+\kappa_\tau)=\frac{1}{2}(\psi^{'}+r^{-1}\sin\psi),  \quad
K=\kappa_{\tau}\kappa_{\nu}=\frac{\psi^{'}\sin\psi}{r}.
\label{eq:s-curvatures}
\end{equation}

Defining $L=\frac{1}{2\kappa}r(W_H)'$, we write the system of first order differential equations governing the problem as \cite{hassinger2017design},

\begin{align}
\begin{split}    
{r}' = \cos{\psi}, \quad  {z}' = \sin{\psi}, \quad r{\psi}' = 2 r H - \sin{\psi}, \quad r {H}' = L + r {C}', \\ \frac{L'}{r} = \frac{p}{k} + \frac{\mathbf{f} \cdot \mathbf{n}}{\kappa} + 2H \left[ \left(H - C \right)^2 + \frac{\lambda}{\kappa} \right] - 2 \left( H - C \right) \left[ H^2 + \left( H - r^{-1} \sin{\psi} \right)^2 \right], \\ {\lambda}' = 2 {\kappa} \left( H - C \right) {C}' -  \mathbf{f} \cdot \mathbf{a_s}.
\label{eq:s-systemofequations}
\end{split}
\end{align}

The applied boundary conditions are

\begin{align}
\begin{split}    
r(0^{+})=0, \quad L(0^{+})=0, \quad \psi(0^{+})=0, \\ z (s_{max})=0, \quad \psi(s_{max})=0, \quad \lambda(s_{max})=\lambda_0.
\label{eq:s-BCs}
\end{split}
\end{align}

In asymmetric coordinates, the manifold area can be expressed in term of arc length \citep{hassinger2017design}

\begin{equation}
a(s)=2\pi \int_0^s r(\xi)d \xi\quad \rightarrow \quad \frac{da}{ds}=2\pi r.
\label{eq:s-area-arclength}
\end{equation}

Eq. (\ref{eq:s-area-arclength}) allows us to convert Eq. (\ref{eq:s-systemofequations}) to an area derivative and prescribe the total area of the membrane.

We non-dimensionalized the system of equations as

\begin{align}
\begin{split}
\zeta=\frac{a}{2 \pi R_0^2},\quad x=\frac{r}{R_0}, \quad y=\frac{y}{R_0}, \quad h=HR_0, \quad c=CR_0, \quad l=LR_0,\\ \lambda^{*}=\frac{\lambda R_0^2}{\kappa_0}, \quad p^*=\frac{pR_0^3}{\kappa_0}, \quad f^*=\frac{fR_0^3}{\kappa_0}, \quad \kappa^*=\frac{\kappa}{\kappa_0},
\label{eq:s-non-dimension}
\end{split}
\end{align}

\noindent where $R_0$ is the radius of the initially circular membrane patch.

Rewriting Eq. (\ref{eq:s-systemofequations}) in terms of Eq. (\ref{eq:s-area-arclength}) and the dimensionless variables Eq. (\ref{eq:s-non-dimension}), we get \cite{hassinger2017design}

\begin{align}
\begin{split}    
x\dot{x} = \cos{\psi}, \quad  x\dot{y} = \sin{\psi}, \quad x^2\dot{\psi} = 2 x h - \sin{\psi}, \quad x^2 \dot{h} = l + x^2 \dot{c}, \\ \dot{l} = \frac{p^*}{\kappa^*} + \frac{\mathbf{f}^* \cdot \mathbf{n}}{\kappa^*} + 2h \left[ \left(h - c \right)^2 + \frac{\lambda^*}{\kappa^*} \right] - 2 \left( h - c \right) \left[ h^2 + \left( h - x^{-1} \sin{\psi} \right)^2 \right], \\ \dot{\lambda^*} = 2 \kappa^* \left( h - c \right) \dot{c} -  \frac{\mathbf{f}^* \cdot \mathbf{a_s}}{x}.
\label{eq:s-area-systemofequations}
\end{split}
\end{align}

\subsubsection{Equation of motion for variable bending and Gaussian moduli}

For nonuniform membrane with variable bending and Gaussian moduli, the governing system of equations in Eq. (\ref{eq:s-systemofequations}) becomes

\begin{align}
\begin{split}    
{r}' = \cos{\psi}, \quad  {z}' = \sin{\psi}, \quad r{\psi}' = 2 r H - \sin{\psi}, \\ r {H}' = L + r {C}'-\frac{r\kappa'}{\kappa} (H-C), \quad {\lambda}' = 2 {\kappa} \left( H - C \right) {C}' -  \mathbf{f} \cdot \mathbf{a_s},\\ \frac{L'}{r} = \frac{p}{k} + \frac{\mathbf{f} \cdot \mathbf{n}}{\kappa} + 2H \left[ \left(H - C \right)^2 + \frac{\lambda}{\kappa} \right] - 2 \left( H - C \right) \left[ H^2 + \left( H - r^{-1} \sin{\psi} \right)^2 \right]\\\-\frac{\kappa'}{\kappa}\frac{L}{r}-\frac{\kappa_G^{''}}{\kappa}\frac{\sin\psi}{r}-\frac{\kappa_G'}{\kappa}\frac{\cos\psi}{r}(2H-\frac{\sin\psi}{r}).
\label{eq:s-variable-systemofequations}
\end{split}
\end{align}

The applied boundary conditions are same as Eq. (\ref{eq:s-BCs}). In addition to Eq. (\ref{eq:s-non-dimension}), we obtain two additional terms.

\begin{align}
\begin{split}
\kappa_G^*=\frac{\kappa_G}{\kappa_0}, \quad \text{and} \quad K^{*}=KR_0^2.
\label{eq:s-variable-non-dimension}
\end{split}
\end{align}

The non-dimensional system of equations in Eq. (\ref{eq:s-area-systemofequations}) becomes,

\begin{align}
\begin{split}    
x\dot{x} = \cos{\psi}, \quad  x\dot{y} = \sin{\psi} \quad x^2\dot{\psi} = 2 x h - \sin{\psi}, \quad x^2 \dot{h} = l + x^2 \dot{c}-x^2\frac{\cdot{\kappa^*}}{\kappa^*}(h-c) \\ \dot{l} = \frac{p^*}{\kappa^*} + \frac{\mathbf{f}^* \cdot \mathbf{n}}{\kappa^*} + 2h \left[ \left(h - c \right)^2 + \frac{\lambda^*}{\kappa^*} \right] - 2 \left( h - c \right) \left[ h^2 + \left( h - x^{-1} \sin{\psi} \right)^2 \right]\\-\frac{\dot{\kappa^*}}{\kappa^*}l-x\frac{\ddot{\kappa}_G^*}{\kappa^*}\sin \psi-\frac{\dot{\kappa}^*_G}{\kappa}\cos\psi(2h-\frac{\sin\psi}{x}), \\ \dot{\lambda^*} = 2 \kappa^* \left( h - c \right) \dot{c}-\dot{\kappa^*}(h-c)^2\dot{c}-\dot{\kappa}^*_GK^* -  \frac{\mathbf{f}^* \cdot \mathbf{a_s}}{x}.
\label{eq:s-variable-area-systemofequations}
\end{split}
\end{align}

\subsubsection{Force balance along the membrane for constant bending and Gaussian moduli}

A general force balance for a surface $\omega$ ``,bounded by a curve $\partial \omega$, is '' (Fig. \ref{fig:schematic}) is

\begin{equation}
\int_{\omega} p\textbf{n} da+\int_{\partial \omega}\tilde{\textbf{f} }dt+\textbf{F}=0,
\label{eq:s-force balance}
\end{equation}

\noindent where $t=r(s)\theta$ is the length along the curve of revolution perimeter, p is the pressure difference across the membrane, $\tilde{\textbf{f}}$ is the traction along the curve of revolution t and $\textbf{\text{F}}$ is a point force applied externally to the membrane. Along any curve like $\partial \omega$ that cuts off the membrane at constant z, the traction is given by \cite{agrawal2009boundary,rangamani2013interaction}

\begin{equation}
\tilde{\textbf{f}} =  \tilde{f}_{\nu} \bm{\nu} + \tilde{f}_{n} \textbf{n}+\tilde{f}_{\tau} \bm{\tau}, 
\label{eq:s-traction:general}
\end{equation}

\noindent where

\begin{align}
\tilde{f}_n = (\tau W_{K})^{'} - 1/2(W_{H})_{,\nu} - (W_{K})_{,\beta} {\tilde{b}}^{\alpha\beta} \nu_{\alpha}, \nonumber\\
\tilde{f}_{\nu} = W + \lambda - \kappa_{\nu}M,\nonumber\\
\tilde{f}_{\tau} = -\tau M,
\label{eq:s-traction:list}
\end{align}

and $\tilde{f}_n$, $\tilde{f}_{\nu}$ and $\tilde{f}_{\tau}$ are force per unit length acting along the normal $\textbf{n}$, tangent $\bm{\nu}$ to the surface and transverse tangent $\textbf{e}_{\theta}$ respectively. In Eq. (\ref{eq:s-traction:list}), $M$ is the bending couple given by

\begin{equation}
M=\frac{1}{2}W_H+\kappa_\nu W_K.
\label{eq:s-bending couple}
\end{equation}

Because $\tau=0$ (no twist) in asymmetric coordinates, the normal and tangential tractions become

\begin{subequations}
\begin{align}
\tilde{f}_n= -\kappa (H'-C') \label{eq:s-traction-normal2}\\ \tilde{f}_{\nu}=\kappa (H-C)(H-C-\psi')+\lambda.
\label{eq:s-traction-tangential2}
\end{align}
\end{subequations}

Projecting Eq. (\ref{eq:s-traction:general}) onto orthogonal bases $\textbf{e}_{r}$ and $\textbf{k}$ gives us the equation for axial and radial tractions \cite{agrawal2009boundary,rangamani2013interaction} 

\begin{subequations}
\begin{align}
\tilde{f}_r=\underbrace{\kappa(H'-C')\sin\psi}_{\text{\shortstack{Curvature gradient \\ contribution}}}+\underbrace{\kappa(H-C)(H-C-\psi')\cos\psi}_{\text{\shortstack{Curvature \\ contribution}}}+\underbrace{\lambda \cos\psi}_{\text{\shortstack{Tension \\ contribution}}}, \label{eq:s-traction-radial}\\ \tilde{f}_z=\underbrace{-\kappa(H'-C')\cos\psi}_{\text{\shortstack{Curvature gradient \\ contribution}}}+\underbrace{\kappa(H-C)(H-C-\psi')\sin\psi}_{\text{\shortstack{Curvature \\ contribution}}}+\underbrace{\lambda \sin\psi} _{\text{\shortstack{Tension \\ contribution}}}.
\label{eq:s-traction-axial}
\end{align}
\end{subequations}

Because $\int_{\partial \omega}dt=2 \pi r$, the applied force in axial direction can be evaluated by substituting Eqs. (\ref{eq:s-traction-radial},\ref{eq:s-traction-axial}) into Eq. (\ref{eq:s-force balance})

\begin{equation}
-\textbf{F}_{z}=\underbrace{2\pi r\left(-\kappa(H'-C')\cos\psi\right)+\kappa(H-C)(H-C-\psi')\sin\psi+\lambda\sin\psi}_{\text{Force due to traction}}+\underbrace{2\pi \int_{0}^{s}p r(\xi)\cos\psi d\xi}_{\text{Force due to pressure}}.
\label{eq:s-axial Force}
\end{equation}

This can be rewritten in terms of tractions as 

\begin{equation}
-\textbf{f}_{z}=\underbrace{\left(-\kappa(H'-C')\cos\psi\right)+\kappa(H-C)(H-C-\psi')\sin\psi+\lambda\sin\psi}_{\text{axial traction}}+\underbrace{ \frac{\int_{0}^{s}p r(\xi)\cos\psi d\xi}{r}}_{\text{Traction due to pressure}},
\label{eq:s-axial Force in traction form}
\end{equation}

where $\textbf{f}_z = \frac{\textbf{F}_{z}}{2 \pi r}$. The energy per unit length $\xi$ for circular deformation at the end of a protein coat can be found by integrating Eq. (\ref{eq:s-traction-radial}) along the perimeter boundary $\partial \omega$,

 \begin{subequations}
\begin{align}
\xi=2\pi r \Big[\underbrace{\kappa(H-C)(H-C-\psi')\cos\psi}_{\text{\shortstack{Curvature\\contribution }}}+\underbrace{\lambda \cos\psi}_{\text{\shortstack{Tension\\contribution }}}+\underbrace{\kappa(H'-C')\sin\psi}_{\text{\shortstack{Curvature gradient\\contribution }}}\Big ].
\label{eq:s-EPUL}
\end{align}
\end{subequations}

\subsubsection{Force balance along the membrane for variable bending and Gaussian moduli}
For a membrane with variable bending moduli, the normal and tangential tractions in Eqs. (\ref{eq:s-traction-normal2}, \ref{eq:s-traction-tangential2}) become

\begin{subequations}
\begin{align}
\tilde{f}_n=-\kappa(H'-C')-\kappa'(H-C)-\frac{\sin \psi}{r}\kappa'_G, \label{eq:s-general-traction-normal}\\ \tilde{f}_{\nu}=\kappa(H-C)(H-C-\psi')+\lambda.
\label{eq:s-general-traction-tangential}
\end{align}
\end{subequations}

 The radial and axial tractions in Eqs. (\ref{eq:s-traction-radial}) and (\ref{eq:s-traction-axial}) can be rewritten for the general case as

\begin{subequations}
\begin{align}
\tilde{f}_r=\kappa(H'-C')\sin\psi+ \kappa(H-C)(H-C-\psi')\cos\psi+\lambda \cos\psi \nonumber\\+\underbrace{\kappa'(H-C)\sin \psi}_{\text{\shortstack{Variable bending \\ modulus}}}+\underbrace{\frac{\sin \psi^2}{r}\kappa'_G}_{\text{\shortstack{Variable Gaussian \\ modulus}}}, \label{eq:s-general-traction-radial}\\ \tilde{f}_z=-\kappa(H'-C')\cos\psi+\kappa(H-C)(H-C-\psi')\sin\psi+\lambda \sin\psi \nonumber\\-\underbrace{\kappa'(H-C)\cos\psi}_{\text{\shortstack{Variable bending \\ modulus}}}-\underbrace{\frac{\sin \psi \cos \psi}{r}\kappa'_G}_{\text{\shortstack{Variable Gaussian \\ modulus}}},
\label{eq:s-general-traction-axial}
\end{align}
\end{subequations}

Similarly, the axial force and energy per unit lengths in Eqs. (\ref{eq:s-axial Force},\ref{eq:s-EPUL}) can be rewritten as

 \begin{subequations}
 \begin{align}
\textbf{F}_{z}=\Big [\underbrace{2\pi r(-\kappa(H'-C')\cos\psi+\kappa(H-C)(H-C-\psi')\sin\psi+\lambda \sin\psi-\kappa'(H-C)\cos\psi-\frac{\sin \psi \cos \psi}{r}\kappa'_G}_{\text{Force due to traction}} \Big ] \nonumber\\+\underbrace{2\pi \int_{0}^{s}p r(\xi)\cos\psi d\xi}_{\text{Force due to pressure}},
\label{eq:s-axial Force-variable moduli}
\end{align}
\end{subequations}

 \begin{subequations}
\begin{align}
\xi=2\pi r \Big[\underbrace{\kappa(H-C)(H-C-\psi')\cos\psi}_{\text{\shortstack{Curvature\\contribution }}}+\underbrace{\lambda \cos\psi}_{\text{\shortstack{Tension\\contribution }}}+\underbrace{\kappa(H'-C')\sin\psi}_{\text{\shortstack{Curvature gradient\\contribution }}}\nonumber\\-\underbrace{\kappa'(H-C)\sin \psi}_{\text{\shortstack{Variable bending\\contribution }}}-\underbrace{\frac{\sin \psi^2}{r}\kappa'_G}_{\text{\shortstack{Variable Gaussian\\contribution }}} \Big ].
\label{eq:s-EPUL-variable moduli}
\end{align}
\end{subequations}

\subsubsection{Asymptotic approximation for small radius}

To ensure continuity at the poles, we use $L=H' = 0$ as a boundary condition in our simulations. However, this boundary condition reduces the expressions for tractions (Eqs. \ref{eq:s-traction-axial}, \ref{eq:s-traction-radial}) to zero at the pole. To avoid this discrepancy, we derive an asymptotic expression for tractions at small arc length. We proceed by assuming that the pole in Eq. (\ref{eq:s-area-systemofequations}) is at $x=0$ and choose a rescaled variable given by

\begin{equation}
    X = \frac{x}{\epsilon},
    \label{eq:s-rescale1}
\end{equation}

Here, $\epsilon$ is a small parameter, so that $X$ is order of one. We can extend this to other small variables in Eq. (\ref{eq:s-area-systemofequations}) near the pole to get

\begin{equation}
    \quad y = y_0+Y\epsilon,\quad \psi = P\epsilon, \quad s = S\epsilon,
    \label{eq:s-rescale}
\end{equation}

\noindent where $Y$, $P$, $S$ are the corresponding rescaled parameters and $y_0$ is membrane height at the pole.

In the simple case with no spontaneous curvature ($C=0$), no external force $\textbf{f}=0$ and no pressure difference $p=0$, we substitute Eqs. (\ref{eq:s-rescale}) and (\ref{eq:s-rescale1}) into Eq. (\ref{eq:s-area-systemofequations}) and use a Taylor expansion to get

\begin{align} 
\dot{X} = 1 - \frac{(P\epsilon)^2}{2}, \quad  \dot{Y} = P\epsilon - \frac{(P\epsilon)^3}{3!}, \quad  \dot{P} = 2 h - \frac{P}{X} + \frac{{\epsilon}^2}{3!}\frac{P^3}{X}, \quad X \dot{h} &= l, \nonumber\\
\dot{l} =  \epsilon^2 2Xh\left[\frac{\lambda^*}{k^*} - \left( h - \frac{P}{X} + \frac{P^3 \epsilon^2}{X3!}\right)^2\right], \nonumber\\
\dot{\lambda^*} = 0.
\label{eq:s-final}
\end{align}

We look for a solutions with form of 

\begin{align} 
h = h^{0} + \epsilon h^{1} +\ord (\epsilon^2), \quad X = X^{0} + \epsilon X^{1}  +\ord (\epsilon^2),\quad Y = Y^{0} + \epsilon Y^{1} +\ord (\epsilon^2), \nonumber\\ l = l^{0} + \epsilon l^{1} +\ord (\epsilon^2), \quad P= P^{0} + \epsilon P^{1} +\ord (\epsilon^2),\quad \lambda^* = \lambda^{*0} + \epsilon \lambda^{*1} +\ord (\epsilon^2).
\label{eq:s-expansion}
\end{align}

The leading order terms in Eq. (\ref{eq:s-expansion}) are

\begin{align} 
\dot{X^{0}} = 1 , \quad  \dot{Y^{0}} = 0, \quad \dot{P^{0}} = 2 h^{0} - \frac{P^{0}}{X^{0}}, \quad  \dot{h^{0}} &= \frac{l^{0}}{X^{0}},\quad
\dot{l^{0}} =  0,\quad
\dot{\lambda}^{*0} = 0.
\label{eq:s-final2}
\end{align}

Integrating the differential equations in Eq. (\ref{eq:s-final2}) , we get

\begin{align} 
X^{0} = S , \quad  Y^{*0} = Y_{0}, \quad P^{0} = S\left(H_{0} + L_{0}\log(S) - \frac{L_{0}}{2}\right),\nonumber\\   h^{0} = L_{0}\log(S) + H_{0},\quad
l^{0} = L_{0},\quad
\lambda^{0} = \lambda_{0}.
\label{eq:s-final3}
\end{align}






\noindent where $Y_{0}$, $H_{0}$ and $L_{0}$, $\lambda_{0}$ are integration constants. We then look at order $\epsilon^{1}$ terms in Eq. (\ref{eq:s-final})

\begin{align} 
\dot{X^{1}} = 0 , \quad  \dot{Y^{1}} = P, \quad \dot{P^{1}} = 2h^{1}+\frac{P^0 X^1}{{X^{0}}^2}, \quad   {X^{0}}\dot{h^{1}}+{X^{1}}\dot{h^{0}} &= l^{1},\quad
\dot{l^{1}} =  0,\quad
\dot{\lambda}^{*1} = 0.
\label{eq:s-final4}
\end{align}

The first order terms are thus given by

\begin{align} 
X^{1} = X_1 , \quad  Y^{1} = P_1S+Y_1, \quad {l^{1}} = L_1,\quad \lambda^{*1}=\lambda_1,\quad  h^{1}= L_1\log(S)+\frac{X_1L_0}{S}+H_1, \nonumber\\ P^{1} = 2S(L_1 \log(S)-L_1+H_1)+X_1L_0\log(S)(\frac{3}{2}+\frac{\log(S)}{2}+\frac{H_0}{L_0}).
\label{eq:s-final5}
\end{align}

Combining the leading order and first order terms and substituting into Eq. (\ref{eq:s-expansion}), our system of variables can be written as 

\begin{align}
X = S + \epsilon X_{1}, \quad Y = Y_{0} + \epsilon(P_1S+Y_1), \quad l = L_{0} + \epsilon L_{1},\quad \lambda^{*} = \lambda_{0} + \epsilon \lambda_{1},\nonumber\\ \quad P = S\left(H_{0} + L_{0}\log(S) - \frac{L_{0}}{2}\right) + \epsilon \Big(2S(L_1 \log(S)-L_1+H_1)+X_1L_0\log(S)(\frac{3}{2}+\frac{\log(S)}{2}+\frac{H_0}{L_0})\Big),\nonumber\\  h = H_{0} + L_{0}\log(S) + \epsilon \Big (L_1\log(S)+\frac{X_1L_0}{S}+H_1 \Big). 
\label{s-combinedscaled}
\end{align}


We are interested in the asymptotic expansion of mean curvature near the pole, which is given by

\begin{equation}
 h = H_{0} + L_{0}\log(S) + \epsilon \Big(L_1\log(S)+\frac{X_1L_0}{S}+H_1 \Big).
\end{equation}

This can be rewritten as 

\begin{align}
 h = H_{0} + L_{0}\log(A + S - A) + \epsilon H_1, \nonumber\\
  h = H_{0} + L_{0}\log(A) + L_{0}\log(1 + \frac{S-A}{A}) + \epsilon \Big (L_1\log(S)+\frac{X_1L_0}{S}+H_1 \Big),
\end{align}

where A is a constant. If $\frac{S-A}{A}$ is small, we can perform a Taylor expansion around $S=A$ to get leading order

\begin{align}
  h = H_{0} + L_{0}\log(A) + L_{0}\left(\frac{S-A}{A} - \frac{1}{2}(\frac{S-A}{A})^2 \ldots \right) \nonumber\\
  h \sim H_{0} + L_{0}\log(A) - L_{0} + L_{0}\left(\frac{S}{A}\right)  \nonumber\\
  h \sim H_{0} + L_{0}\left(\log(A) - 1 + \frac{S}{A}\right) \nonumber\\
  h \sim H_{0} + L_{0}\log(A) - L_{0}+L_{0}\left( \frac{s}{A\epsilon}\right)\nonumber\\
  h \sim C_{1} + C_{2}s,
\end{align}

\noindent where $C_{1}$ and $C_{2}$ are constants. This shows that the mean curvature can be approximated as a linear solution near the pole for $S \sim A$ or $s \sim A\epsilon$. In our image analysis, inaccuracies near the pole begin at orders of magnitude of $10^{-2}$. At this range, we can approximate a linear solution for mean curvature. 

Similarly, we consider an asymptotic expansion for $\psi$ near the pole at leading order 

\begin{equation}
 P = S\left(H_{0} + L_{0}\log(S) - \frac{L_{0}}{2}\right),
 \end{equation}
 
 which can be rewritten as

 \begin{align}
 \psi = s\left(H_{0} + L_{0}\log(s) - L_{0}\epsilon -\frac{L_{0}}{2}\right) \rightarrow
 \psi = s\left(D_{1} + D_{2}\log(s)\right),
 \end{align}

 \noindent where $D_{1}$ and $D_{2}$ are constants. We can now substitute the approximation for mean curvature and $\psi$ near the pole into Eq. (\ref{eq:s-traction-radial}) and (\ref{eq:s-traction-axial}) to get

\begin{subequations}
\begin{align}
\tilde{f}_r \sim -\kappa(C_{1} + C_{2}s-C)(C_{1} + C_{2}s-C -D_{2} -  D_{1} - D_{2}\log(s))-\lambda,\\
\tilde{f}_z \sim -\kappa(C_{2}-C').
\label{eq:s-frfzpresimple}
\end{align}
\end{subequations}

Using $\log(s) = \log(s+A-A) = \log(A) + \log(1 + \frac{s-A}{A})$ and expanding around $s \sim A$, Eq. (\ref{eq:s-frfzpresimple}) can be simplified to 

 \begin{subequations}
 \begin{align}
 \tilde{f}_r \sim -\kappa(F_{1}s^2 + F_{2}s+F_{3})-\lambda,\\
\tilde{f}_z \sim -\kappa(C_{2}),
\label{eq:s-asymptotic}
\end{align}
\end{subequations}

\noindent where $F_{1}$, $F_{2}$ are constants.  We can thus approximate radial traction as quadratic in arc length near the pole, while axial traction can be correspondingly approximated as constant. In this work, we choose to start the asymptotic solution at the local minimum of mean curvature near the pole, which is $\epsilon$ $\sim$ 0.1. Figures 2-5 in the main text are plotted using this relation. 

\newpage

\section{Table of notation}

\begin{table}[h!]
\begin{center}
\caption{Notation used in the model}
\begin{tabular} {l l l}
\hline\hline
Notation &  Description & Units\\ [0.5ex]
\hline
$E$ & Strain energy  & pN $\cdot$ nm \\
$\gamma$ & Lagrange multiplier for incompressibilty constrain   & pN$/$nm \\
$p$ & Pressure difference across the membrane  & pN$/$nm$^2$ \\
$C$ & Spontaneous curvature & nm$^{-1}$ \\
$\theta^{\alpha}$ & Parameters describing the surface \\ 
$W$ & Local energy per unit area &  pN/nm\\
${\bf r}$ & Position vector & \\
${\bf n}$ & Normal to the membrane surface &  unit vector\\
${\bm{ \nu}}$ & Tangent to the membrane surface in direction of increasing arc length &  unit vector\\
${\bm{ \tau}}$ & Rightward normal in direction of revolution  &  unit vector\\
$\textbf{a}_\alpha$ & Basis vectors describing the tangent plane\\
$\lambda$ & Membrane tension, $-(W+\gamma)$  & pN/nm\\
$H$ & Mean curvature of the membrane & nm$^{-1}$\\
$K$ & Gaussian curvature of the membrane & nm$^{-2}$\\
$\kappa$ & Bending modulus (rigidity)  & pN $\cdot$ nm\\
$\kappa_G$ & Gaussian modulus & pN $\cdot$ nm\\
$s$ & Arc length  &  nm\\
$\theta$ & Azimuthal angle  & \\
$\psi$ &  Angle between $ \textbf{e}_{r}$  and  $\textbf{a}_{s}$ & \\
$r$ & Radial distance  & nm \\
$z$ & Elevation from base plane & nm \\
$\textbf{e}_{r} (\theta)$ & Radial basis vector  & unit vector \\
$\textbf{e}_{\theta} $ & Azimuthal basis vector  & unit vector \\
$\textbf{k} $ & Altitudinal basis vector  & unit vector \\
$\textbf{F}$ & External force  &  pN \\
${\bf f}$ & Applied force per unit area &  pN $/$nm$^2$\\
$\kappa_{\tau}$ & Transverse curvature& nm$^{-1}$\\
$\kappa_{\nu}$ & Tangential curvature& nm$^{-1}$\\
$\tau$ & Surface twist & nm$^{-1}$\\
$\tilde{\textbf{f}}$& Traction (force per unit length)& pN$/$nm\\
$\tilde{f_r}$ & Component of traction in radial direction & pN$/$nm\\
$\tilde{f_z}$ & Component of traction in axial direction & pN$/$nm\\
$\tilde{f_n}$ & Component of traction in normal direction & pN$/$nm\\
$\tilde{f_{\nu}}$ & Component of traction in transverse direction & pN$/$nm\\
$\tilde{F_z}$ & Calculated force in axial direction & pN\\
$\xi$ & Energy per unit length & pN\\
$M$ & Bending couple & pN $\cdot$ nm\\
$t$ & Arc length around curve of revolution  & nm\\
$a$ & Membrane area  & $\mathrm{nm^2}$\\
$V$ & Confined volume by membrane area  & $\mathrm{nm^3}$\\
$s_{max}$ & Maximum arc length at the base  & nm\\
$R_0$ & Patch radius  & nm\\
$\kappa_0$ & Bending rigidity of bare membrane  & pN $\cdot$ nm\\
\hline
\end{tabular}
\end{center}
\label{table:notation}
\end{table}

\begin{table}[h!]
\begin{center}
\caption{Notation used in the model}
\begin{tabular} {l l l}
\hline\hline
Notation &  Description & Units\\ [0.5ex]
\hline
$\lambda_0$ & Surface tension at boundary  & pN$/$nm\\
$L$ & Shape equation variable  & $\mathrm{nm^{-1}}$\\
$x$ & Dimensionless radial distance  \\
$y$ & Dimensionless height  \\
$h$ & Dimensionless mean curvature  \\
$c$ & Dimensionless spontaneous curvature  \\
$l$ & Dimensionless L  \\
$\lambda^*$ & Dimensionless surface tension  \\
$p^*$ & Dimensionless pressure  \\
$f^*$ & Dimensionless force per unit area  \\
$\kappa^*$ & Dimensionless bending modulus\\
$\kappa_G^*$ & Dimensionless Gaussian modulus  \\
$K^*$ & Dimensionless Gaussian curvature  \\
$\zeta$ & Dimensionless area  \\
$A$ & Area of spontaneous curvature field  & $\mathrm{nm^2}$\\
$\zeta_{\mathrm{force}}$ & Area of the applied force  & $\mathrm{nm^2}$\\
$\epsilon$ & Small parameter \\
$X$ & Rescaled parameter for x\\
$Y$ & Rescaled parameter for y\\
$P$ & Rescaled parameter for $\psi$\\
\hline
\end{tabular}
\end{center}
\label{table:notation}
\end{table}

\section{Analysis of experimental images}

\begin{figure}[h!]
\centering
\includegraphics[width=5 in]{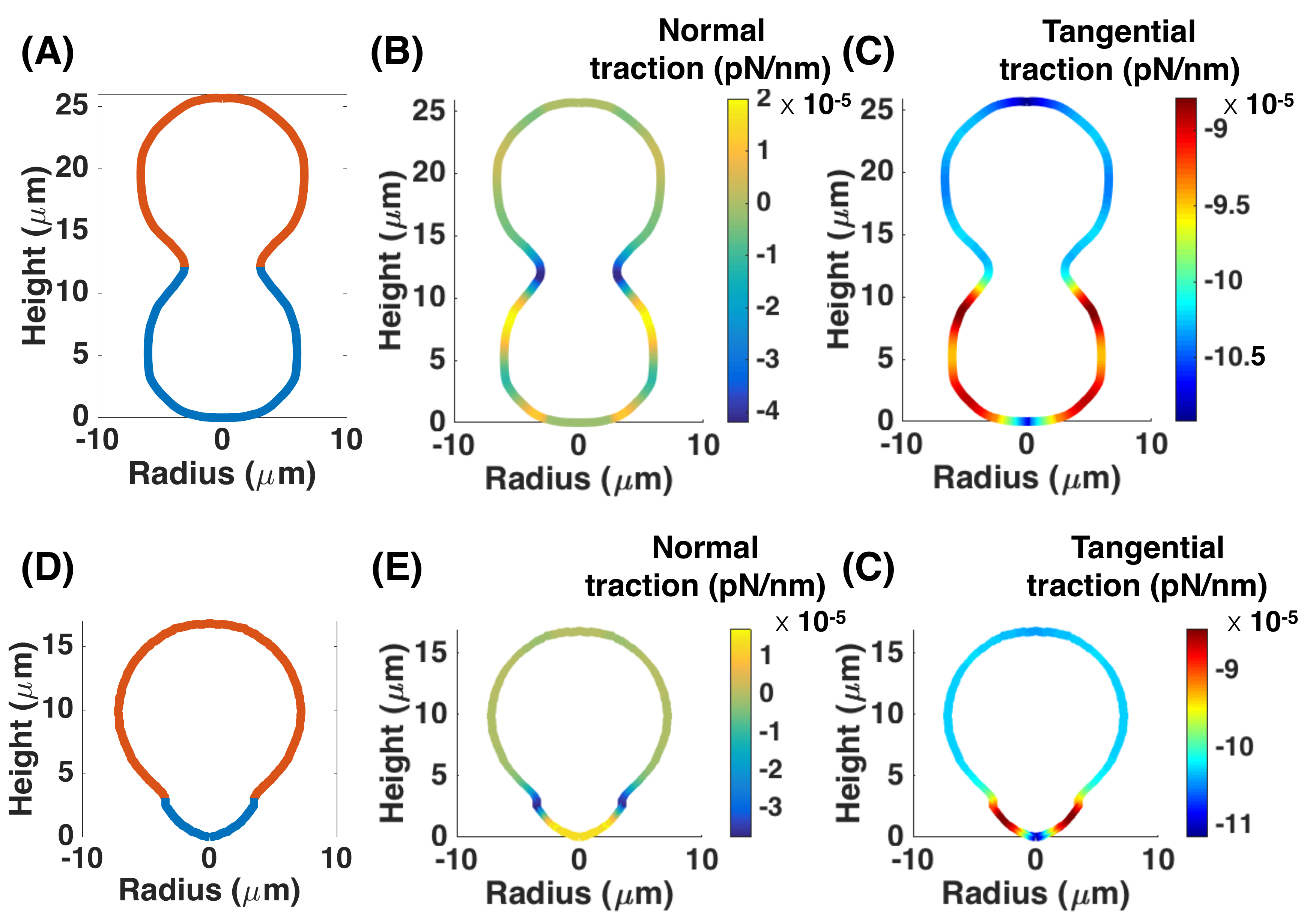}
\caption{Normal and tangential traction distributions along vesicles with fluid phase coexistence shown in (A) and (D). Red is lipid disordered phase (L$_d$) and blue is lipid ordered phase (L$_o$). Parameters used were p = $2.8 \,\times 10^{-2} \,\mathrm{N/m^2}$, surface tension of disordered phase $\lambda_d = -1.03 \,\times 10^{-4} \,\mathrm{mN/m}$, surface tension of ordered phase  $\lambda_o = -0.91 \,\times 10^{-4} \,\mathrm{mN/m}$, bending modulus of disordered phase $\kappa_d = 10^{-19} \,\mathrm{J}$, ratio of bending modulii $\kappa_o/\kappa_d = 5$ and absolute difference in Gaussian moduli $\Delta \kappa_G = 3.6 \times 10^{-19} \,\mathrm{J}$ \cite{baumgart2005membrane}. Normal and tangential tractions were calculated using Eqs. (\ref{eq:s-general-traction-radial}, \ref{eq:s-general-traction-tangential}). (B), (E) Normal traction distribution along corresponding vesicle shapes. Large negative traction observed at the interface, which is also the neck. Calculating the energy per unit length (Eq. \ref{eq:s-EPUL-variable moduli}) at this point predicts a line tension of 0.7 and 0.8 pN respectively - very close to the experimentally determined value of 0.67 pN. (C), (F) Tangential traction distribution along the corresponding vesicle shapes. L$_d$ phase has larger magnitude of tangential traction than L$_o$ phase, consistent with Fig. \ref{fig:baumgart} in main text. Gradient in a tangential traction observed at the interface.}
\label{Baumgart A}
\end{figure}

We extract `x' and `y' data from images obtained from previously published works \cite{baumgart2005membrane,baumgart2003imaging}. Images are converted to greyscale and analyzed to obtain an outline of the membrane using `ImageJ.' We then extract coordinates and import them into MATLAB where we compute the angle $\psi$, mean curvature H, and tractions at every point based on Eq. (\ref{eq:s-general-traction-radial}) and (\ref{eq:s-general-traction-axial}). Here, we note that x = 0 at both poles. To obtain the asymptotic solution for another pole at the base, we rescale our variables as

\begin{equation}
    \quad x = X\epsilon, \quad y =  Y\epsilon,\quad \psi = P\epsilon, \quad s = s_{max} + S\epsilon,
    \label{s-rescale3}
\end{equation}

where $s_{max}$ is arc length at the base pole. This rescaling gives us the same asymptotic solution as Eq. (\ref{eq:s-asymptotic}). Other parameters like pressure, bending modulus and surface tension of L$_d$ and L$_o$ phase, difference in Gaussian modulus were taken from Baumgart et al. \cite{baumgart2005membrane} and implemented using a hyperbolic tangent function. Fig. \ref{fig:baumgart} in main text showed the traction distributions along one vesicle shape. Fig. \ref{Baumgart A}  plots tractions along two other experimental vesicle shapes \cite{baumgart2005membrane}. We see that the defining characteristics are similar in that the normal traction is large and negative at the neck/interface of the two domains and tangential traction shows a large gradient at the same point. Calculating the energy per unit length (Eq. \ref{eq:s-EPUL-variable moduli}) for the two vesicle shapes shown in Fig. \ref{Baumgart A} (A) and (D) gives values of 0.7 and 0.8 pN respectively, which is close to experimentally determined value.

\section{Additional tether and bud formation simulations}

\subsection{Tubes pulled against surface tension} 



\begin{figure}[h!]
\centering
\includegraphics[width=5 in]{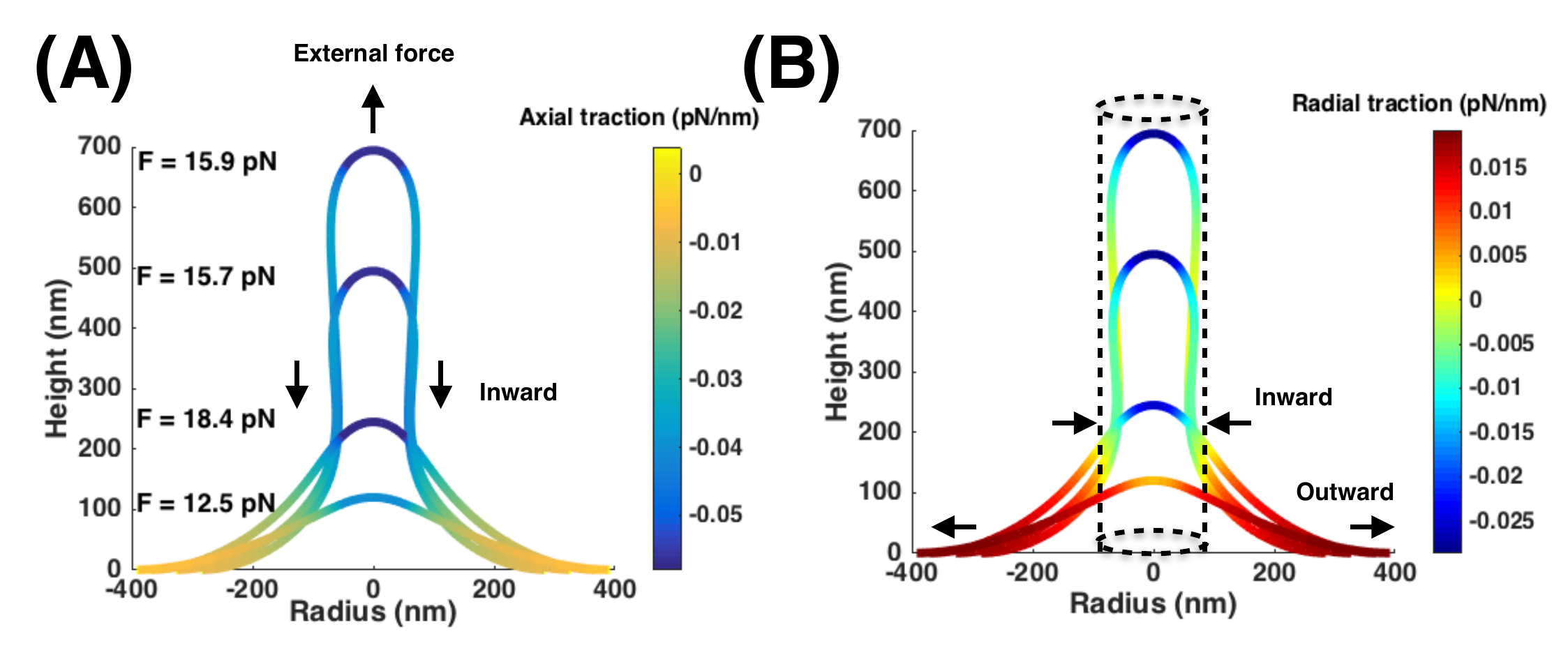}
\caption{Axial and radial traction (Eqs.( \ref{eq:s-traction-axial}, \ref{eq:s-traction-radial})) distribution plotted along shapes in Fig. \ref{fig:tubes_simulation}. (A) Axial traction distribution. Large negative value is observed at the pole. Membrane curves away from the applied force leading to large axial traction response. (B) Radial traction distribution. Dotted cylinder is the stable cylindrical geometry. Values of radial traction within cylinder are negative and those outside are positive. Radial traction can be interpreted as a membrane response in the radial direction to tether formation where the membrane tries to bend inward along the tether and push outward along the base.}
\centering
\label{Derenyi_axial_radial}
\end{figure}

Motivated by Derenyi et al \cite {derenyi2002formation}, we recreate membrane tube pulling by applying a point axial load to a circular patch of membrane shown in Fig. 2 of the main text. We set up the simulation to calculate the axial force needed to achieve a membrane tube of specified height and map the normal and tangential traction along the membrane. We used a bending modulus of $320 \mathrm{pN \cdot nm}$, surface tension of $0.02 \mathrm{pN/nm}$, and applied force over $\%1.5625$ of the membrane area (approximating a point force). Fig. \ref{Derenyi_axial_radial} plots the axial and radial components of the traction along the same equilibrium shapes. To compare the traction force to external force, we use Eq. (\ref{eq:s-force balance}) and modify it to be a function of non dimensional membrane area $\zeta$ (Eq.(\ref{eq:s-non-dimension})). For a simulation with no pressure and in axisymmetric coordinates ($dt = 2\pi r(s)$), Eq. (\ref{eq:s-force balance}) simplifies to


\begin{equation}
2\pi r(s)\tilde{\textbf{f} }+\int_{\partial \omega}\textbf{f} 2\pi R_0^2d\alpha=0,
\label{eq:s-fbalance}
\end{equation}

where $\textbf{f}$ is force per unit area applied externally to the membrane. We use a hyperbolic tangent function to define the applied force, which is given by 

\begin{equation}
\textbf{f} = \frac{\textbf{F}}{2\pi R_0^2 \zeta_{\mathrm{force}}} \frac{\tanh(g(\zeta-\zeta_{\mathrm{force}}))}{2},
\label{eq:s-f_area}
\end{equation}

where $\zeta_{\mathrm{force}}$ is the non-dimensional area of the applied force, $g=20$ is a constant and $\textbf{F}$ is applied force. We can substitute Eq. (\ref{eq:s-f_area}) into Eq.(\ref{eq:s-fbalance}) to get 

\begin{subequations}
\begin{align}
2\pi r(s) \tilde{\textbf{f} } = -\int_{0}^{\zeta} \frac{\textbf{F}}{\zeta F} d\zeta, \quad \zeta < \zeta_{\mathrm{force}} \label{fbalance_1}\\
2\pi r(s) \tilde{\textbf{f} } = -\textbf{F} , \quad \zeta \geq \zeta_{\mathrm{force}}.
\label{eq:s-fbalance_2}
\end{align}
\end{subequations}

Eqs. (\ref{eq:s-fbalance_2}, \ref{eq:s-fbalance_2}) relate external force to axial membrane traction at every point along the membrane. 

\subsection{Tubes pulled against pressure} \label{lab:lowpressure}

In Fig. \ref{fig:tubes_simulation}, we set $p = 0$ and $\lambda = 0.02\ \mathrm{pN/nm}$. However, pressure plays an important role in tether formation and cannot be ignored \citep{dmitrieff2015membrane}. We investigated the role pressure plays during tether formation by finding an equivalent pressure to surface tension that can produce a tube of similar radius. To do this, we first define a natural length scale for the system, $R_0$, by the expected equilibrium radius of a membrane tube obtained by minimization of the free energy of the membrane \cite{derenyi2002formation}.

In absence of pressure, external force, spontaneous curvature and Gaussian modulus, we can write the free energy Eq. (\ref{eq:s-energy density}) of the membrane as

\begin{equation} \label{eq:s-freeE}
E = \int_{\omega} (\kappa H^2+\lambda) \mathrm{d}a.
\end{equation}

For a tube of length $L$ and radius $R$, the free energy, ignoring the mean curvature of the cap ($H=\frac{1}{2R}$), can be written as 

\begin{equation} \label{eq:s-freeEtube}
\mathcal{W}_{tube} = \left( \frac{\kappa}{4 R^2} + \lambda \right) 2 \pi R L.
\end{equation}

The balance between the surface tension, which acts to reduce the radius, and the bending rigidity sets the equilibrium radius $R_0$. Taking $\partial \mathcal{W}_{tube} / \partial R = 0$ we obtain

\begin{eqnarray} \label{eq:s-R_0f0}
R_0 \equiv \frac{1}{2} \sqrt{\frac{\kappa}{ \lambda}}.
\end{eqnarray}

We can perform a similar analysis with pressure replacing surface tension. The free energy of the membrane Eq. (\ref{eq:s-freeE}) can be rewritten as

\begin{equation} \label{eq:s-freeEp}
E = \int_{\omega} \kappa H^2 \mathrm{d}a + p V.
\end{equation}

Again for a tube of length $L$ and radius $R$, the free energy can be written as 

\begin{equation} \label{eq:s-freeEtubep}
\mathcal{W}_{tube} = \left( \frac{\kappa}{4 R^2}\right) 2 \pi R L +p \pi R^2 L.
\end{equation}

Here, the balance between pressure, which acts to reduce the radius, and the bending rigidity sets the equilibrium radius $R_0$. Taking $\partial \mathcal{W}_{tube} / \partial R = 0$ we obtain

\begin{eqnarray} \label{eq:s-R_0f0p}
R_0 \equiv \sqrt[3]{\frac{\kappa}{4p}}.
\end{eqnarray}

Comparing Eq. (\ref{eq:s-R_0f0p}) and Eq. (\ref{eq:s-R_0f0}), we can find an equivalent pressure to the surface tension needed for achieving a tube of radius $R_0$, 

\begin{align}
\sqrt[3]{\frac{\kappa}{4p}} =  \frac{1}{2} \sqrt{\frac{\kappa}{ \lambda}},\nonumber\\
p = \frac{2 \lambda \sqrt{\lambda}}{\sqrt{\kappa}}.
\label{eq:s-pandlam}
\end{align}

\begin{figure}[h!]
\centering
\includegraphics[width=6 in]{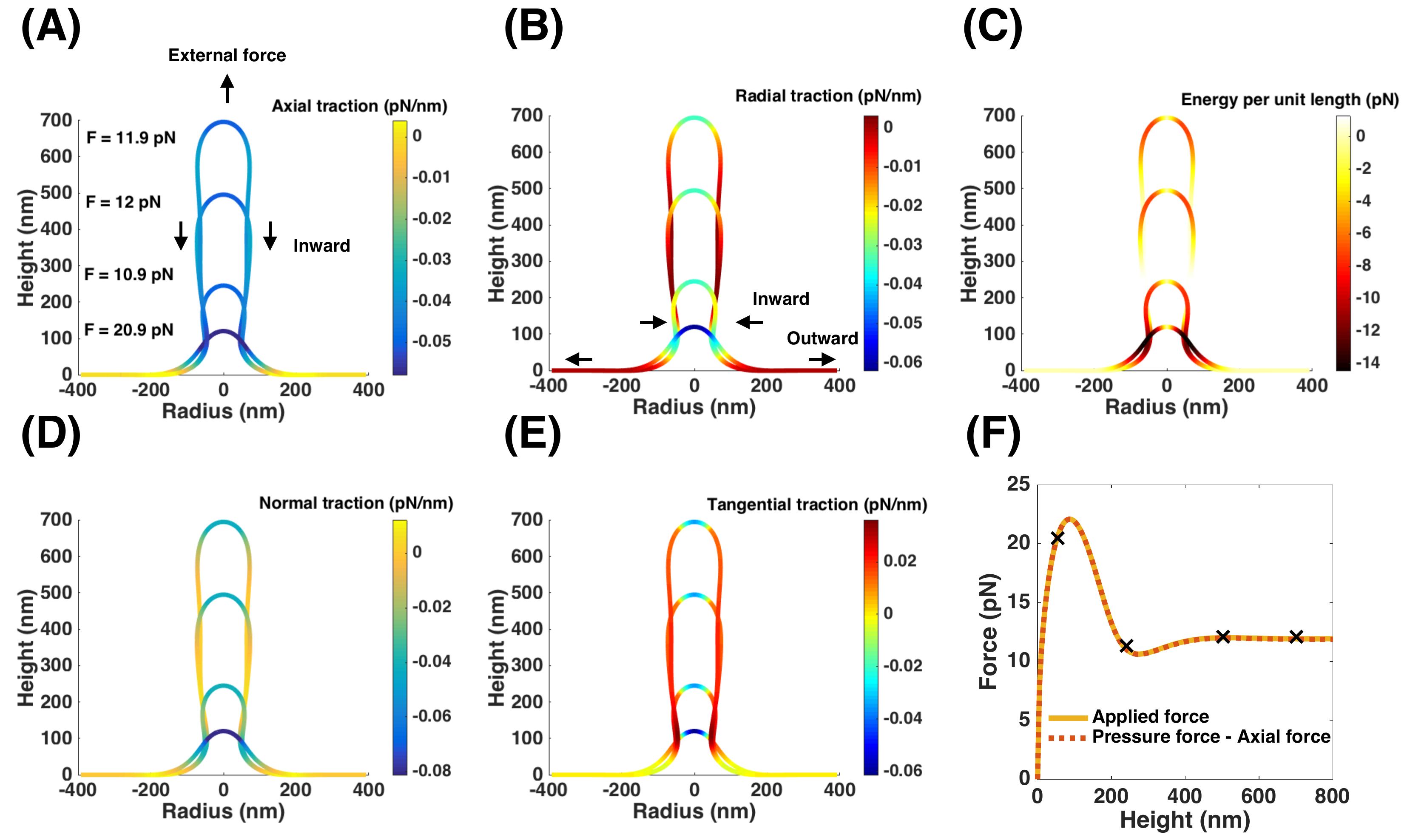}
\caption{Tether pulling simulation for pressure of $0.3 \,\mathrm{kPa}$, bending modulus $320 \,\mathrm{pN \cdot nm}$, no surface tension at the boundary ($\lambda=0$), and a point force. We seek to simulate an equivalent membrane tube to Fig. \ref{fig:tubes_simulation} by pulling a tether out against pressure instead of surface tension. (A) Axial traction distribution along the tether. Axial traction is uniform with a maximum at the pole, and becomes negligible at the base. Axial traction at pole is particularly large for initial shapes as the tether is pulled out since the membrane is trying to pull out against a pressure. (B) Radial traction distribution - We see a negative value at the neck and a positive value at the base. This can be interpreted as a membrane response to tether formation in radial direction  where the membrane likes to bend in at the neck and likes to push out at the base. (C) Energy per unit length Eq. (\ref{eq:EPUL}) plotted along the shapes. Large value observed at the neck - predicting an `effective' line tension of 11 pN to form a tether of height 700 nm. (D) Normal traction is negative at the neck and over the area of applied force and is negligble elsewhere. Large values are observed at the pole for initial shapes, similar to axial traction. (E) Tangential traction changes sign from negative at the pole to positive along the tether and finally becomes zero at the base. Large gradient in surface tension observed at the neck (F) Applied force plotted alongside difference between calculated pressure and axial force (Eq. \ref{eq:s-axial Force-variable moduli}). Exact match observed, verifying  the accuracy of the result.}
\centering
\label{tubewithpress_nolam}
\end{figure}

Eq. (\ref{eq:s-pandlam}) gives an equivalent pressure $p = 0.3 \,\mathrm{kPa}$ for a surface tension of $0.02 \,\mathrm{pN/nm}$. We perform the tether pulling simulation for this value of pressure, such that the pressure acts inward for every non-zero height. Surface tension is set to zero at the base. Fig. \ref{tubewithpress_nolam} A and B map the axial and radial traction along the tether and Fig. \ref{tubewithpress_nolam} D and E plot the corresponding normal and tangential components. The traction distributions show similar behaviour to Fig. \ref{fig:tubes_simulation} in the main text. Using Eq. (\ref{eq:s-axial Force-variable moduli}), the applied force matches the difference between pressure force in axial direction and the force due to axial traction (Fig. \ref{tubewithpress_nolam} F). Panel C plots the energy per unit length (Eq. \ref{eq:s-EPUL-variable moduli}) - it shows similar behavior to Fig. \ref{fig:tubes_simulation}D in the main text.


\subsection{Tubes pulled against pressure and surface tension}

\begin{figure}[h!]
\centering
\includegraphics[width=5.6 in]{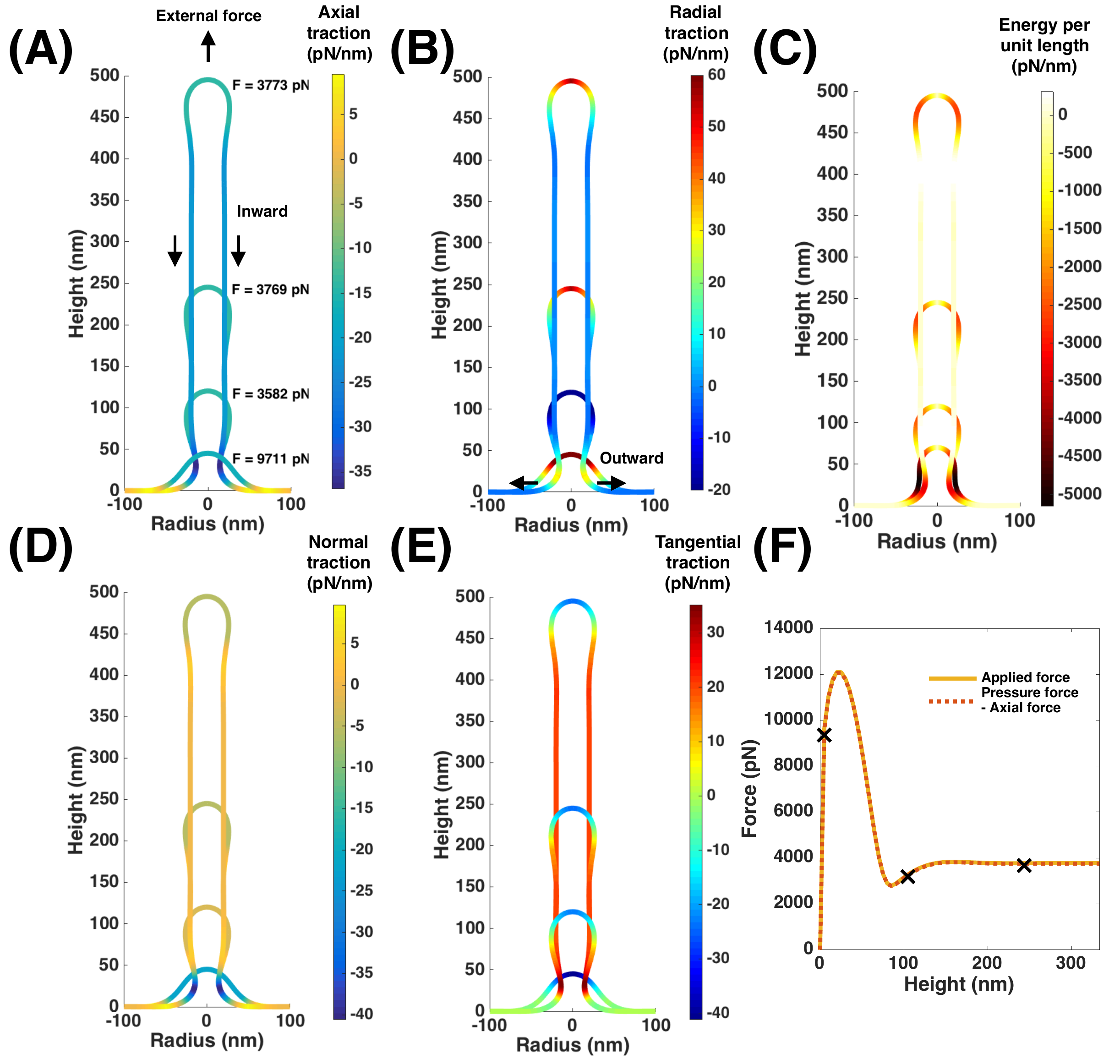}
\caption{Tether pulling simulation for pressure of $1 \,\mathrm{MPa}$, surface tension $0.02 \,\mathrm{pN/nm}$, bending modulus $32000 \,\mathrm{pN \cdot nm}$ and point force. The membrane tube is narrower and requires a much larger external force to counter the effect of pressure and tension. (A) Axial traction distribution along the tether. Traction is uniform and negative along the cylindrical part representing a uniform membrane response, and becomes negligible at the base. The small positive region at the base is where pressure suddenly drops to zero and is the membrane response to the sudden lack of pressure. (B) Radial traction distribution - Large positive value at the neck, indicating the tendency of membrane to avoid forming a narrow neck and push out. (C) Energy per unit length Eq. (\ref{eq:EPUL}) along the shapes - large value observed at the neck predicting an `effective' line tension of 3300 pN. (D) Normal traction distribution. Zero normal traction along the tube and large value observed at the neck. (E) Tangential traction is almost constant and positive along the tube. There are two sign change in tangential traction, (\textit{i}) at the end of applied force, (\textit{ii}) at the neck where the tether attaches to base. Gradient in surface tension observed at the neck and near the pole where the surface tension changes sign
(F) Force match plotted vs height of the tether (Eq. ( \ref{eq:s-axial Force-variable moduli}). Applied force can be matched to the difference between pressure force and axial force.}
\centering
\label{tubewithpress}
\end{figure}

Typically, yeast endocytic buds experience a very large pressure in the order of $1 \,\mathrm{MPa}$ \cite{basu2014role, dmitrieff2015membrane}. In Fig. \ref{tubewithpress}, we perform the tether pulling simulation for pressure $1 \,\mathrm{MPa}$, surface tension $0.02 \,\mathrm{pN/nm}$ and bending modulus of $32000 \,\mathrm{pN \cdot nm}$, suggested by \cite{dmitrieff2015membrane}. Fig.  \ref{tubewithpress} A and B map the axial and radial tractions for four membrane shapes as the tether is pulled out. Because of the large pressure, the radius of the tether is very small. A consequence of the small radius is a positive radial traction at the neck, where the membrane wants to push out. Axial and radial traction are both constant over cylindrical parts of the tether. Energy per unit length, seen in Fig \ref{tubewithpress} C shows large negative values at the neck and near the pole, similar to cases before. Fig \ref{tubewithpress} D and E plot normal and tangential traction distributions along the membrane are also the same as former results, but differs in magnitude due to larger bending modulus, tractions being almost two order of magnitudes larger. In Fig. (\ref{tubewithpress} F) the external force is plotted vs the height of the tether and matches the difference between pressure force and axial force (Eq.  \ref{eq:s-axial Force-variable moduli}). In the presence of pressure, a much larger force is required to pull out the tube. The maximum force is almost 600 times larger than the case without pressure.

\begin{figure}[t!]
\centering
\includegraphics[width=3in]{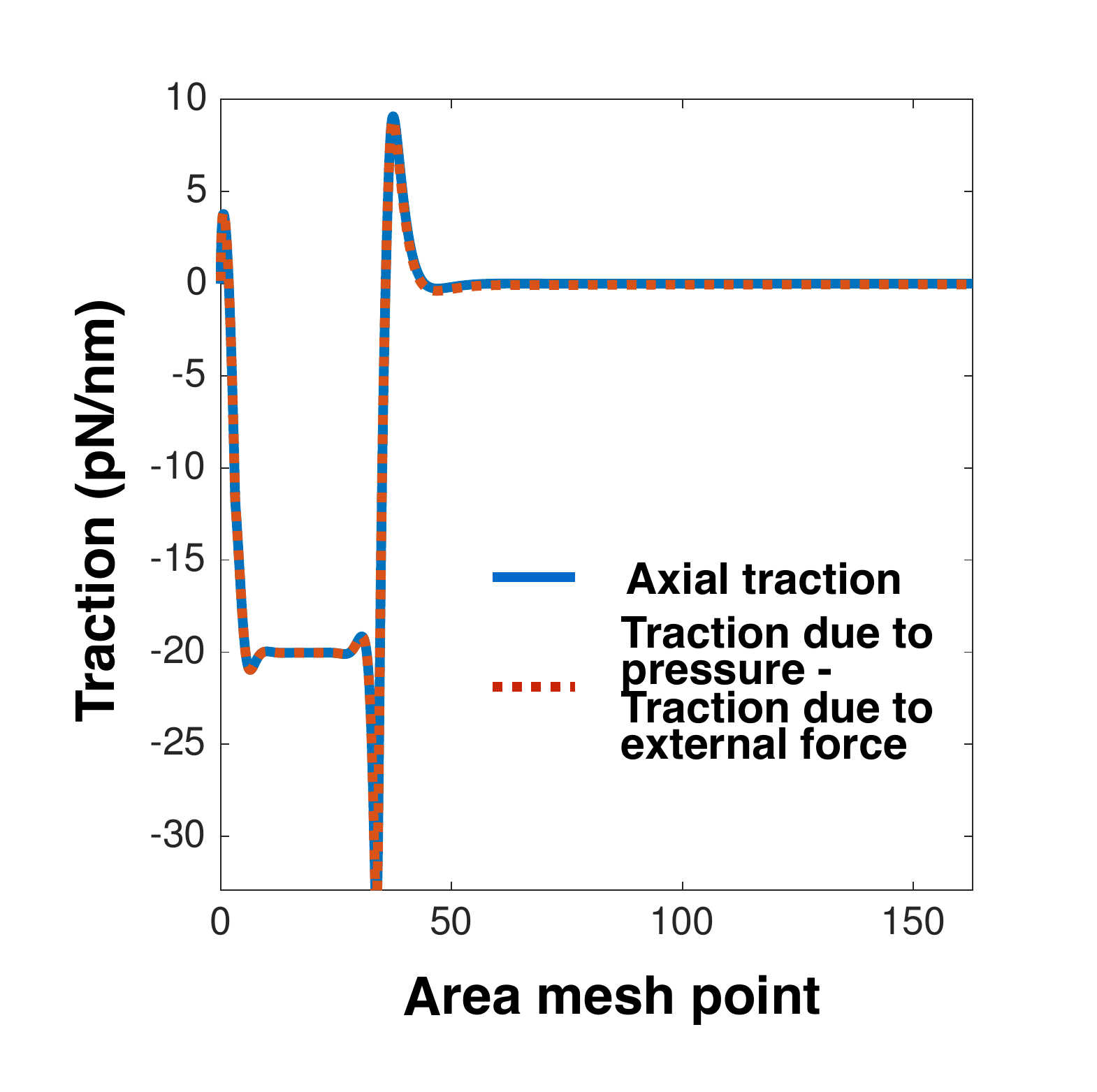}
\caption{Tractions plotted along a membrane tether of height $500 \,\mathrm{nm}$ for a tube pulling simulation with pressure of $1 \,\mathrm{MPa}$, bending modulus $32000 \,\mathrm{pN \cdot nm}$ and surface tension $0.02 \,\mathrm{pN/nm}$. Axial traction can be matched at every point to tractions due to pressure and external force (Eq. (\ref{eq:s-axial Force in traction form})).}
\centering
\label{tractionmatch}
\end{figure}

Using Eq. (\ref{eq:s-axial Force in traction form}), we can also match the tractions at every point on the membrane to traction due to pressure and traction due to external force. Fig. \ref{tractionmatch} shows tractions plotted along area mesh points for the tube simulation with $p=1 \,\mathrm{MPa}$ and $\kappa=32000 \,\mathrm{pN \cdot nm}$ at a height of $500 \,\mathrm{nm}$.

\subsection{Axial and radial tractions in bud formation}

\begin{figure}[h!]
\centering
\includegraphics[width=5in]{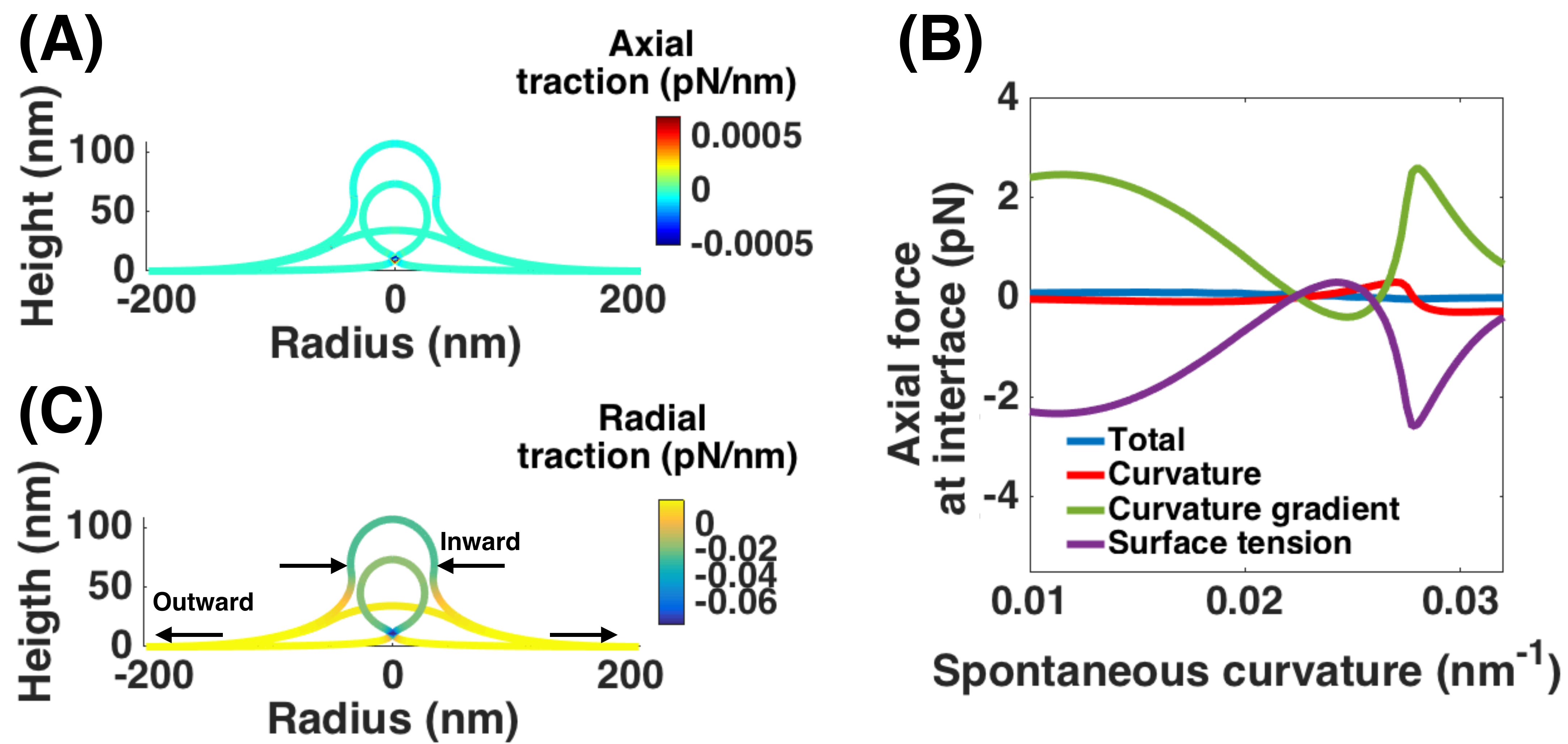}
\caption{Bud formation from a flat membrane for increasing spontaneous curvature and a constant area of spontaneous curvature field $\mathrm{A}=10,053 \,\mathrm{nm}^2$. The spontaneous curvature magnitude is increasing from $C = 0$ to $C = 0.034 \,\mathrm{nm^{-1}}$, the bending modulus is $\kappa=320 \,\mathrm{pN \cdot nm}$ and surface tension at the edge is $\lambda=0.02 \,\mathrm{pN/nm}$. Axial traction does not play any role in invagination. (A) Axial traction along the membrane  is negligible for all shapes. (B) Axial force at the interface is almost zero. Terms due to tension and curvature gradient cancel each other and force due to curvature is automatically zero. (C) Radial traction distribution for three different shapes. Large negative radial traction at the neck can help membrane scission.}
\centering
\label{fig:bud-axial-radial}
\end{figure}

Axial and radial tractions for heterogeneous bud simulations, Fig. (\ref{fig:bud_simulation}) of the main text, are shown in Fig (\ref{fig:bud-axial-radial}). Axial traction along the membrane is negligible in all the stages of bud formation (Fig. \ref{fig:bud-axial-radial}A). Axial force due to traction Eq. (\ref{eq:s-axial Force}) depends on three different terms, curvature, curvature gradient and surface tension. Calculated axial force at the interface is zero because tension term cancels out the force due to curvature gradient and the force associated with curvature is zero by itself (Fig. \ref{fig:bud-axial-radial} B). This means that neck formation is purely regulated by radial stresses (Fig. \ref{fig:bud-axial-radial}C). For small deformations, the radial traction is positive throughout, which shows that the membrane works to oppose the deformation. However, with the formation of U shaped caps, radial traction changes sign and acts inward, representing the membrane tendency to form small necks. \\\\\\\\\\\\\\\\\\\\\\\\\\\\\

\subsection{Surface tension at the boundary regulates line tension at interface}

\begin{figure}[h!]
\centering
\includegraphics[width=5in]{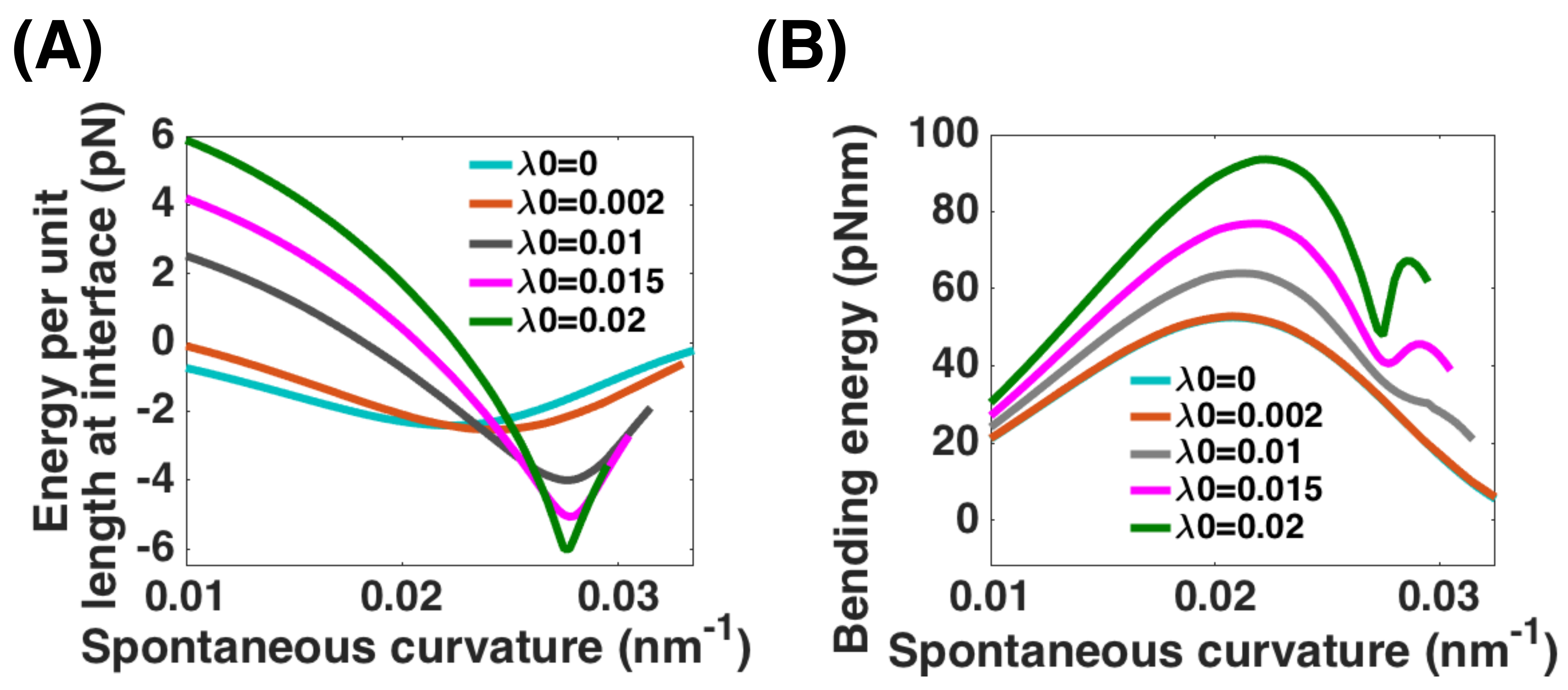}
\caption{Increasing the energy per unit length at interface with increasing surface tension at the boundary Eq. \ref{eq:s-EPUL}. (A) Line tension variation at the interface vs the applied spontaneous curvature field for different values of surface tension at the edge. For large values of surface tension, the energy per unit length has a sign change from positive to negative with increasing spontaneous curvature. (B) Bending energy (Eq. (\ref{eq:s-energy density2})) for different surface tension boundary conditions plotted versus the spontaneous curvature. There is an increase in bending energy cost with increase in surface tension at the boundary.}
\label{fig:tensioncomp}
\end{figure}

Line tension at the interface depends on the surface tension value at the boundary (Fig. \ref{fig:tensioncomp}). For zero surface tension at the boundary, curvature gradient is the only dominant term in Eq .(\ref{eq:s-EPUL}) and line tension is always negative (Fig. \ref{fig:tensioncomp} A).
With increasing value of surface tension at the boundary, the line tension behavior can be classified in two different regimes; (1) tension dominant (2) curvature gradient dominant. In both regimes, the magnitude of line tension is larger for higher values of surface tension at the boundary. The larger line tension at the interface can be associated with an increase in bending energy, calculated by Eq. (\ref{eq:s-energy density2})  (Fig. \ref{fig:tensioncomp} B).

\subsection{Bud formation in heterogeneous membrane with negative surface tension at boundary}

In Fig. (\ref{fig:bud_simulation}), surface tension at the boundary is set to $\lambda=0.02 \,\mathrm{pN/nm}$, which is close to its biological value  \cite{stachowiak2013cost,karpova2000role}. This means that at the edge where the membrane connects to reservoir, there is a tensile stress. However, it could be possible that the boundary applies compressive stress to the domain of interest. Here, we repeat the simulation for a heterogeneous membrane with negative surface tension value at the boundary $\lambda=-0.02 \,\mathrm{pN/nm}$.



\begin{figure}[h!]
\centering
\includegraphics[width=5 in]{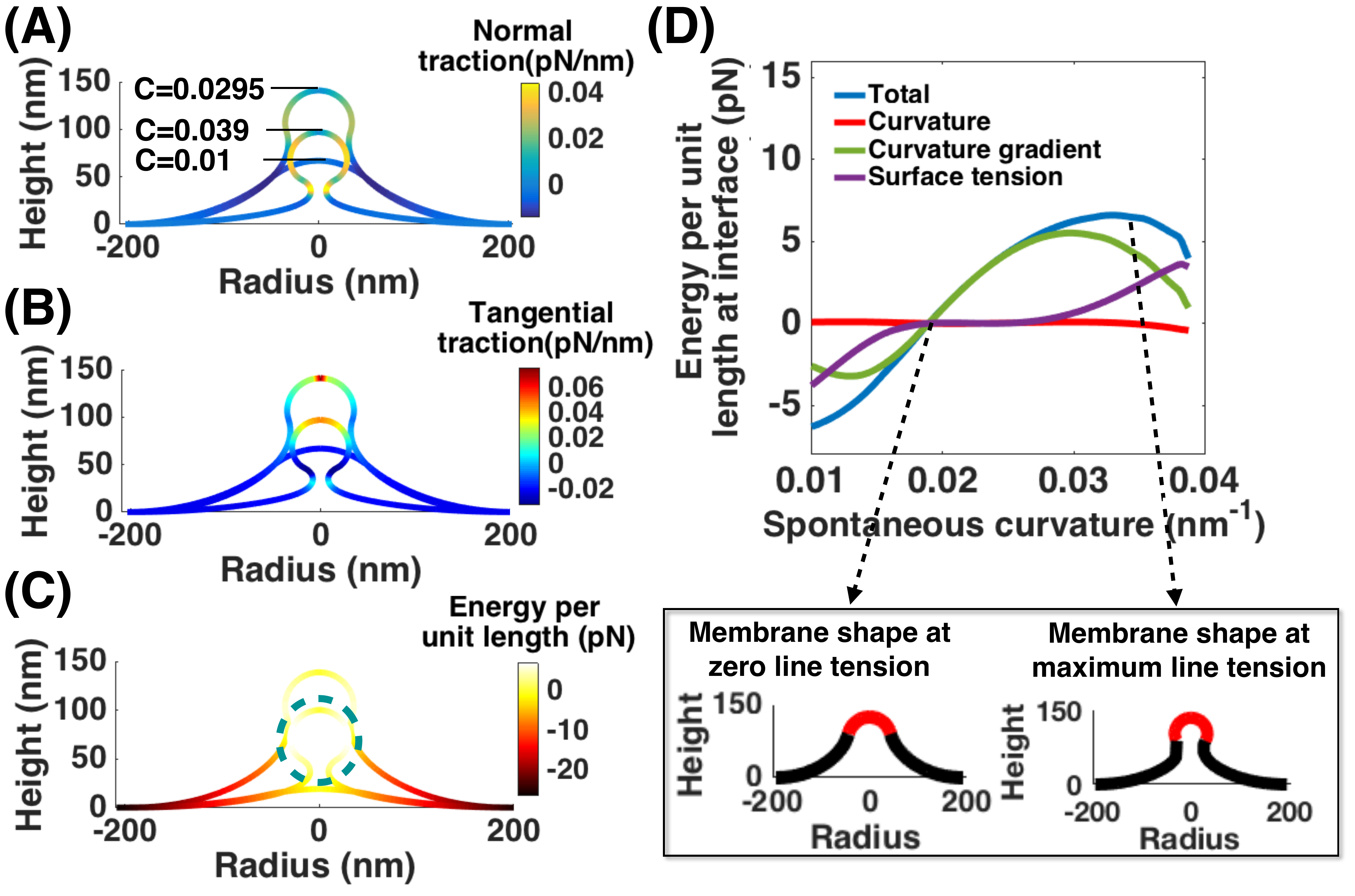}
\caption{ Bud formation with increasing spontaneous curvature magnitude ($C = 0$ to $C = 0.039 \,\mathrm{nm^{-1}}$) and negative surface tension at the edge $\lambda = -0.02 \,\mathrm{pN/nm}$. The area of the spontaneous curvature field and bending modulus are the same before. Here, larger spontaneous curvature is required to form a bud because the negative surface tension opposes the bending force. (A) Positive normal traction along the cap and bud represents the `membrane resistance against deformation'. (B) Tangential traction distribution along the membrane for three different shapes. In contrast to Fig. \ref{fig:bud_simulation}, tangential traction at the neck jumps from negative value to positive indicating the membrane tendency to open the vesicle (C) Positive energy per unit length inside the equilibrium vesicle -dashed circle- shows how `effective' line tension opposes bending deformation. (D) Energy per unit length always has the same trend as curvature gradient. The subplots show the membrane configuration at zero and maximum line tension, with red lines representing the protein coat coverage for each shape.}
\label{fig:budnegativetension}
\end{figure}

Fig. \ref{fig:budnegativetension} shows how a bud forms from an initially flat membrane by  increasing spontaneous curvature magnitude from $C = 0.01$ to $C = 0.039 \,\mathrm{nm^{-1}}$. Normal traction along the bud is positive, showing the membrane resistance against deformation (Fig. \ref{fig:budnegativetension}A). Here, larger spontaneous curvature is required to form a bud compared  to Fig. (\ref{fig:bud_simulation}) due to the unfavorable gradient in tangential traction at the neck- the sharp rise from negative before the neck to positive value after the neck (Fig. \ref{fig:budnegativetension}B). The energy per unit length inside the equilibrium vesicle -dashed circle- is positive indicating that the negative surface tension at the boundary is an unfavorable condition for bud formation (Fig. \ref{fig:budnegativetension} C). The initial negative energy per unit length  represents membrane tendency for buckling (Fig. \ref{fig:budnegativetension}D).  However, for large values of spontaneous curvature, positive energy per unit length is required to balance the negative surface pressure resulting in stable intermediate shapes  (open and U shaped buds). Indeed, the positive line tension is essential to get a smooth-shape evolution from a flat membrane to a closed bud. 

\end{document}